\documentclass{article}
\usepackage{arxiv}
\usepackage[utf8]{inputenc}
\usepackage[T1]{fontenc}
\usepackage{authblk}
\usepackage{amsmath,amssymb,amsfonts,amsthm}
\usepackage{booktabs}
\usepackage{graphicx}
\usepackage{listings}
\usepackage{xcolor}
\usepackage{natbib}
\usepackage{hyperref}
\usepackage{url}
\usepackage{cleveref}

\newtheorem{Definition}{Definition}
\newtheorem{Lemma}{Lemma}
\newtheorem{Corollary}{Corollary}
\newtheorem{Proposition}{Proposition}
\newtheorem{Theorem}{Theorem}
\theoremstyle{remark}\newtheorem*{Remark}{Remark}

\title{Self-Referential Introspection in Large Language Models: the Critical Threshold for Recursive Self-Improvement}
\date{}

\newbox{\orcidicon}\sbox{\orcidicon}{\includegraphics[height=8pt]{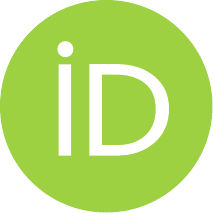}}
\author[1,2]{\href{https://orcid.org/0000-0001-7402-7482}{\usebox{\orcidicon}\hspace{0.5mm}Jiang Zhang}\thanks{Corresponding author: \texttt{zhangjiang@bnu.edu.cn}}}
\author[2]{\href{https://orcid.org/0009-0007-9062-0383}{\usebox{\orcidicon}\hspace{0.5mm}Bing Yuan}\thanks{\texttt{yuanbing@swarma.org}}}
\author[2]{\href{https://orcid.org/0009-0006-9084-8356}{\usebox{\orcidicon}\hspace{0.5mm}Qian Zhang}\thanks{\texttt{zhangqian@swarma.org}}}
\affil[1]{School of Systems Science, Beijing Normal University}
\affil[2]{Swarma Research}
\renewcommand{\shorttitle}{Self-Referential Introspection in Large Language Models}
\makeatletter
\renewcommand{\@toptitlebar}{%
  \noindent\includegraphics[width=0.25\textwidth]{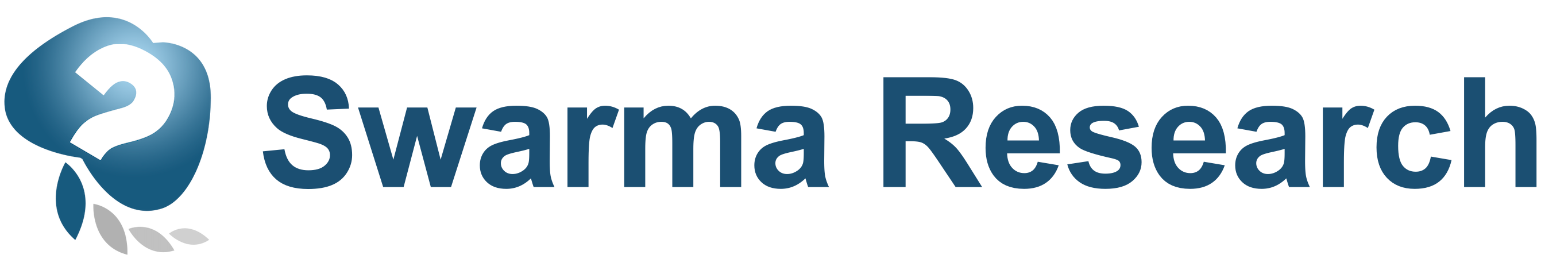}\par
  \vskip 0.15in
  \hrule height 2\p@
  \vskip 0.25in
  \vskip -\parskip%
}
\makeatother
\hypersetup{
hidelinks,
pdftitle={Self-Referential Introspection in Large Language Models: the Critical Threshold for Recursive Self-Improvement},
pdfauthor={Jiang Zhang, Bing Yuan and Qian Zhang},
pdfsubject={Recursive self-improvement, self-reference, and introspection in large language models},
pdfkeywords={self-reference; large language models; recursive self-improvement; introspection; Kleene's second recursion theorem; von Neumann's complexity threshold; metacognition; AI self-evolution}
}

\begin{document}
\maketitle
\begin{abstract}
\texorpdfstring{
The pursuit of self-evolving AI raises a critical question: when is autonomous self-improvement sustainable rather than degenerative? Drawing an analogy to von Neumann's complexity threshold for self-reproducing automata, we argue that sustainable recursive self-improvement in large language models (LLMs) requires a functional analogue: introspection---the system's capacity to simulate its own operations and target modifications. Grounded in Kleene's Second Recursion Theorem, we construct such introspective self-improvement programs and prove their key properties: completeness of self-modification, necessity of the reflective architecture, undecidability of improvement in general, and equivalence with Schmidhuber's G\"odel machine under a rewrite-equivalence notion, which transfers the global optimality guarantee. An empirical review, organized around these functional criteria, suggests that current LLMs exhibit only \emph{quasi-introspection}. The available evidence does not establish complete introspection in the formal sense developed here, while pointing to several candidate structural bottlenecks, including incomplete self-access, feedforward processing, and limited computational depth. We outline architectural paths toward the threshold and discuss the safety implications of crossing it.
}{The pursuit of self-evolving AI raises a critical question: when is autonomous self-improvement sustainable rather than degenerative? Drawing an analogy to von Neumann's complexity threshold for self-reproducing automata, we argue that sustainable recursive self-improvement in large language models (LLMs) requires a functional analogue: introspection---the system's capacity to simulate its own operations and target modifications. Grounded in Kleene's Second Recursion Theorem, we construct such introspective self-improvement programs and prove their key properties: completeness of self-modification, necessity of the reflective architecture, undecidability of improvement in general, and equivalence with Schmidhuber's G\"odel machine under a rewrite-equivalence notion, which transfers the global optimality guarantee. An empirical review, organized around these functional criteria, suggests that current LLMs exhibit only quasi-introspection. The available evidence does not establish complete introspection in the formal sense developed here, while pointing to several candidate structural bottlenecks, including incomplete self-access, feedforward processing, and limited computational depth. We outline architectural paths toward the threshold and discuss the safety implications of crossing it.}
\end{abstract}
\section{Introduction: The Bet on Self-Evolving AI and Introspection}

The past two years have witnessed a decisive strategic shift across the AI industry. Major laboratories---OpenAI, Google DeepMind, Anthropic, Meta, Alibaba, and others---are no longer content with building static foundation models that must be retrained by human engineers. Instead, they are investing heavily in \emph{self-evolving} AI systems: agents that autonomously refine their own prompts, strategies, tool usage, memory structures, and even underlying parameters through iterative interaction with the environment \citep{peng2026,gao2026,tao2026}.

This ambition is not new. The concept of \emph{recursive self-improvement} (RSI)---a system that not only improves itself but improves its capacity to improve itself---has been a central idea in the theory of artificial general intelligence since at least Good's \citeyearpar{good1965} ``intelligence explosion'' hypothesis and Yudkowsky's \citeyearpar{yudkowsky2007} Seed AI framework. What is new is that large language models (LLMs), for the first time, appear to provide a plausible substrate for such systems. Recent works have demonstrated LLMs that refine their own reasoning through self-reflection \citep{shinn2023,madaan2023}, optimize their own prompts \citep{fernando2023}, generate and evaluate their own training data \citep{huang2023,yuan2024}, and even modify their own source code at runtime \citep{yin2024,zhang2026}.

Yet a sobering empirical pattern has emerged: most autonomous self-improvement loops exhibit rapid asymptotic saturation. For instance, the Self-Refine paradigm plateaus after a mere few iterations \citep{madaan2023}. This systemic limitation is clarified by \citet{huang2024large}, who demonstrated that absent external ground-truth feedback, contemporary LLMs lack the intrinsic corrective capacity to rectify their own reasoning errors, frequently suffering performance degradation following self-revision. Consequently, even the most sophisticated self-evolving frameworks---such as the Darwin G\"odel Machine \citep{zhang2026}---remain strictly tethered to external benchmarks and human-engineered evaluation criteria to sustain behavioral progress.

Another critical obstacle to sustained self-improvement in LLMs is \emph{model collapse}---a degenerative process wherein successive generations of generative models are progressively polluted by their own synthetic outputs \citep{shumailov2024nature}. \citet{dohmatob2025} further demonstrated that even a minuscule fraction of data pollution can trigger this collapse, meaning that scaling the training set ceases to yield performance gains. Scaling this effect globally, as AI-generated text permeates the web and shifts human reliance, the broader knowledge economy risks tipping into a steady state of "knowledge collapse," where foundational general knowledge gradually vanishes despite the availability of high-quality personalized advice \citep{knowledge_collapse2026}. Consequently, a fundamental question emerges: \emph{Is there an absolute barrier to sustained recursive self-improvement, and if so, what is its formal nature?}

We propose that this bottleneck is deeply intertwined with von Neumann's concept of a \textbf{complexity threshold}''---a term we use here to encapsulate his original phrasing, the critical size of complication'' \citep{burks1966}. Through a comparative analysis of man-made automata and biological organisms, von Neumann investigated how a machine might replicate itself. He observed that \textit{there is a minimum number of parts below which complication is degenerative, in the sense that if one automaton makes another the second is less complex than the first, but above which it is possible for an automaton to construct other automata of equal or higher complexity}'' \citep{burks1966}. While admitting the difficulty of strictly defining complication''---abstractly describing it as ``\textit{\ldots the potentiality to do things}''---he maintained that this minimum threshold is dictated by the capacity for self-reproduction. Crucially, he deduced that a self-reproducing automaton requires a dual architecture: the functional automaton itself and a tape containing its own description \citep{burks1966}. 
Kolmogorov complexity\cite{kolmogorov1965} and related algorithmic-information measures\cite{chaitin1966} may provide formal proxies for descriptional complexity in the discrete setting. We do not assume, however, that these measures define a universal scalar complexity or threshold for LLMs.

Fundamentally, self-reproduction is a physical manifestation of self-reference \citep{hofstadter1979geb}, and the theoretical existence of such self-reproducing automata follows directly from Kleene's second recursion theorem \citep{kleene1952metamathematics,cutland1980,rogers1967}. Thus, the complexity threshold that prevents degenerative decay in recursive replication is fundamentally rooted in the mathematics of self-reference.

This paper draws a structural analogy between von Neumann's
self-reproducing automata and contemporary self-improving AI systems. We
formulate the hypothesis that an analogous complexity threshold may govern
LLM self-evolution. Because no scalar metric for an LLM's ``complexity'' is
currently available, we characterize this threshold functionally: it
concerns whether a system can form, access, and use a sufficiently complete
description of the computational unit undergoing improvement. Crossing this
threshold may make sustained autonomous recursive self-improvement possible,
whereas below it, improvement may remain partial, externally scaffolded, or
vulnerable to degradation. The precise outcome depends on the system's
evaluation criteria, modification mechanism, and architectural organization.

We argue that \emph{introspection}---the capacity of a system to simulate and
evaluate its own operational behavior---is a candidate necessary condition
for complete autonomous recursive self-improvement in the formal setting
developed here. Using Quine-like techniques, we construct an introspection
program that simulates its own source code within an embedded virtual machine
\citep{cutland1980}. This construction establishes the computable existence
of a self-referential program and provides a formal basis for analyzing
self-modification and self-improvement architectures, including related
proposals such as the G\"odel machine \citep{schmidhuber2003}. It does not,
by itself, establish that current LLMs implement such a mechanism.

The Quine-based construction has a reflective, two-part organization
comprising an active execution body and a generative description of that
body. For complete self-modification, the computational unit undergoing
improvement must be able to recover and modify an adequate description of
its own relevant structure. If the unit is a bare LLM, this requirement
extends to its parameters and the mechanisms responsible for their
execution, although the description need not be a literal, uncompressed
copy of every weight. This requirement creates a structural challenge for
modern LLMs: current systems exhibit emergent, quasi-introspective behaviors,
such as metacognitive self-monitoring, but existing evidence does not show
that they possess complete self-access or self-modification authority in the
formal sense developed here. We therefore treat the claim that current LLMs
have not crossed the introspection threshold as an architectural assessment,
not as a direct consequence of Kleene's theorem alone.

\begin{figure}[htbp]
  \centering
  \includegraphics[width=1.0\linewidth]{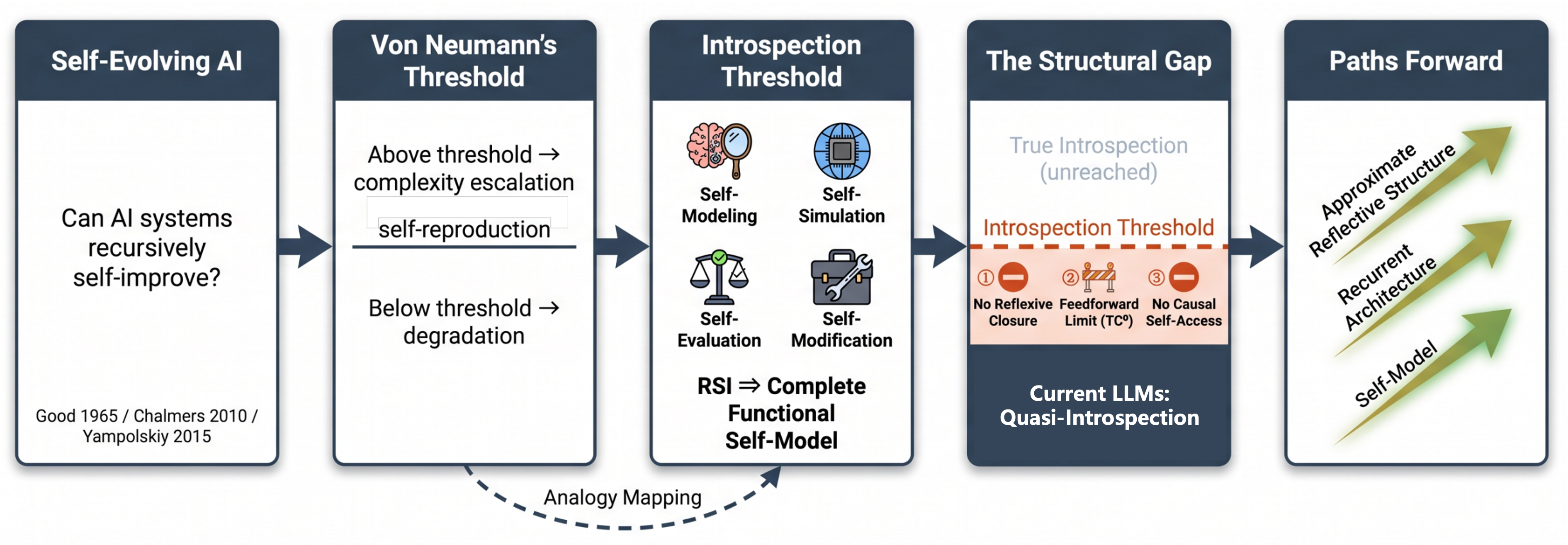}
  \caption[Conceptual framework of recursive self-improvement and introspection]{
  \textbf{Recursive self-improvement and introspection in current LLMs.}
  The diagram relates recursive self-improvement to complexity and introspection thresholds, summarizes evidence for quasi-introspective capabilities in current LLMs, and distinguishes proposed structural limitations from corresponding research directions. Dashed lines denote the proposed introspection threshold.
  }
  \label{fig:outline}
\end{figure}

This paper is organized as follows (see Figure~\ref{fig:outline}):

\begin{enumerate}
  \item We review the literature on recursive self-improvement and
  self-evolving LLMs and AI systems, classify existing approaches, and
  identify their limitations with respect to autonomous and sustained
  self-improvement (Section~\ref{sec:self-improved}).

  \item We revisit von Neumann's complexity threshold and develop the formal
  theory of self-reference, including Quine constructions and Kleene's
  Second Recursion Theorem. We use this framework to construct introspective
  self-improvement programs and establish their formal properties under
  explicit computability and architectural assumptions (Section~\ref{sec:self-reference} and
  Appendix~\ref{sec_appendixA}). Their relation to Schmidhuber's G\"odel machine is analyzed formally in Appendices~\ref{sec:gmalign}--\ref{sec:gmequiv}.

  \item We survey empirical studies of self-modeling, self-simulation,
  self-evaluation, and self-modification in contemporary LLMs. We
  characterize current systems as exhibiting \emph{quasi-introspection},
  while finding no evidence that they satisfy complete introspection in the
  formal sense developed in Section~\ref{sec:self-reference} and \ref{sec:quasi-intro_llm}.

  \item We examine the structural gap between current LLM architectures and
  the formal requirements of introspective self-improvement, discuss possible
  paths toward the introspection threshold, and consider its implications
  for consciousness, recursive self-improvement, AI safety, and institutional
  governance (Sections~\ref{sec:struct-gap-possible-path}--\ref{sec:discussion}). Section~\ref{sec:conclusion} concludes with open problems.
\end{enumerate}

\section{Self-Improved AI Systems}

\label{sec:self-improved}

\subsection{From the Singularity to Recursive Self-Improvement}
\label{sec:singularity}

The hypothesis that machine intelligence could surpass human capabilities and trigger runaway growth dates back to \citet{good1965}, who observed that an ultraintelligent machine'' could design increasingly superior systems, culminating in what he termed an \emph{intelligence explosion}. This positive-feedback loop—wherein a system enhances its own capacity for design—constitutes the conceptual bedrock of \textbf{recursive self-improvement (RSI)}. Although formal terminology emerged decades later, \citet{yudkowsky2001gisai,yudkowsky2007} codified RSI within the framework of seed AI,'' defining it as self-improvement that expands the system’s \emph{capacity to improve} rather than merely optimizing its task performance. Concurrently, \citet{schmidhuber2003}'s G\"odel Machine provided this concept with its strictest mathematical realization: an agent capable of rewriting any part of its own code, provided it can formally prove the modification beneficial.

This conceptual evolution yields a clear chain of causal dependencies: a \emph{singularity} is the hypothesized end state; an \emph{intelligence explosion} is the process of realization; \emph{RSI} is the driving mechanism; and—as argued from Section~\ref{sec:self-reference} onward—RSI itself fundamentally hinges on \emph{self-reference}. By tracing this causal chain to its bedrock, we establish the core contribution of this paper: rather than debating whether an intelligence explosion is imminent or inevitable, we investigate the precise formal capacity a system must possess for self-improvement to compound recursively rather than degenerate.

\subsection{A Classification of Self-Improving Systems}
\label{sec:classification}

To formalize \emph{which} capacity undergoes modification, \citet{yampolskiy2015} proposes a three-level meta-taxonomy:
\begin{itemize}
    \item \textbf{Level~1 (Self-modification)} alters system code without necessarily improving performance (e.g., polymorphic viruses);
    \item \textbf{Level~2 (Weak RSI)} optimizes performance within a fixed algorithmic framework, inevitably yielding diminishing returns (e.g., neural architecture search);
    \item \textbf{Level~3 (Strong RSI)} improves the underlying capability to improve, initiating a potentially unbounded feedback loop. 
\end{itemize}

The boundary between Weak and Strong RSI hinges on whether the improvement mechanism itself has improved. Schmidhuber’s G\"odel Machine instantiates this transition precisely: upon proving a single target theorem, all subsequent meta-levels collapse into a uniform layer\citep{schmidhuber2003}. To date, no deployed system has convincingly achieved Level~3.

Mapping existing models onto this spectrum allows us to survey concrete self-evolving LLMs and agents along two functional axes: the specific architectural layer modified, and the mechanism utilized to verify improvements.

\subsection{Self-Evolving LLMs and Agents: Which Layer, and How Verified?}
\label{sec:layers}
Virtually all contemporary LLMs and agents qualify as Level~2 (Weak RSI) self-improving systems, as they fundamentally rely on machine learning algorithms to optimize task performance. Crucially, by leveraging in-context learning or test-time adaptation, these models can execute self-improvement dynamically during inference. In this section, we review the literature through the critical lens of the self-modification layer.

The \emph{layer} of self-modification defines the specific locus within a system's computational stack where architectural changes are applied. Existing frameworks span four escalating layers: (i) \textbf{prompt/context}, (ii) \textbf{memory/skill}, (iii) \textbf{weight}, and (iv) \textbf{scaffolding-code}. Empirically, Self-Refine \citep{madaan2023} and Reflexion \citep{shinn2023} operate at the prompt/context layer through textual self-critique. SkillRL \citep{xia2026} and AlphaApollo \citep{zhou2025} inhabit the memory/skill layer by accumulating reusable libraries or tools. Moving deeper, GENOME \citep{zhang2025} targets the weight layer by applying evolutionary operators directly to LoRA parameters. At the most fundamental scaffolding-code layer, STOP \citep{zelikman2024}, G\"odel Agent \citep{yin2024}, and the Darwin G\"odel Machine (DGM) \citep{zhang2026} permit systems to rewrite their own orchestration code or manage agent archives. Conversely, RISE \citep{qu2024} cuts across these layers by fine-tuning models to iteratively revise failed execution attempts.

Beyond the modification locus, a self-improving system must rigorously evaluate the validity of its proposed structural changes. Contemporary literature exhibits a definitive consensus: pragmatic deployments have abandoned the uncomputable verification guarantees of Schmidhuber’s original G\"{o}del Machine \citep{schmidhuber2003} in favor of localized statistical approximations. For instance, the Huxley G\"{o}del Machine (HGM) leverages \emph{Clade-Metaproductivity} (CMP) to estimate the long-term fitness of evolutionary lineages within bounded tree searches. Alternatively, the Red Queen G\"{o}del Machine (RQGM) \citep{iacob2026rqgm} employs \emph{controlled utility evolution}, co-evolving evaluators and agents within localized epochs to stabilize non-stationary optimization signals. Sliding further down the formal spectrum, heuristic frameworks like STOP \citep{zelikman2024} bypass internal verification entirely, validating programmatic modifications solely via empirical performance on external benchmarks. Rather than duplicating comprehensive existing surveys \citep{peng2026,gao2026,tao2026}, our analysis untangles the fundamental structural constraints that compel these diverse architectures to externalize their verification signals.

\subsection{Limitations}
\label{sec:limitations}
The empirical limitations of contemporary self-improvement split into internal evaluative constraints and orthogonal distributional hazards.

The primary bottleneck resides in the unreliability of the internal feedback loop. Textual self-refinement routinely saturates within few iterations \citep{madaan2023}, and models routinely fail to self-correct even when errors are explicitly localized \citep{qu2024}. Furthermore, internal evaluation can be actively counterproductive, inducing ``metacognitive hallucinations'' where flawed reflection reinforces incorrect beliefs \citep{lu2025}. This untrusted feedback restricts systems to shallow self-modifications, as the model cannot reliably project the downstream utility of its own structural changes.

Conversely, model collapse introduces an independent bottleneck inherent to the recursive data substrate. This pathology is triggered when AI-generated outputs inadvertently pollute the original training substrate. As successive generations of models are trained on data increasingly contaminated by historical synthetic outputs, the system progressively erodes distributional diversity and collapses functional modes—a degenerative process that remains robust to scale and is highly sensitive to even minimal synthetic fractions \citep{shumailov2024nature,dohmatob2025}. We re-interpret model collapse not merely as a data-quality issue, but as a collapse of effective information entropy successive generational training on synthetic outputs reduces the Shannon entropy / Kolmogorov complexity of the training distribution, analogous to a statistical mechanical mixing process approaching a degenerate fixed point. Because this degradation occurs independently of architectural integrity—affecting even a theoretically perfect self-model—it represents an orthogonal explanation for capability saturation. 

Whether observed test-time optimization reflects genuine strategy or mere capability elicitation remains contested. Nonetheless, the empirical consensus is clear: current self-improvement consistently saturates, internal verification is fragile, and self-modification remains surface-level. This systemic failure indicates that current architectures remain bounded below a fundamental capability threshold. Bridging this gap requires moving beyond incremental engineering toward a rigorous formalization of how a program operates upon itself, positioning \emph{self-reference} as our foundational mathematical bedrock.

\section{Self-reference: from self-reproducing to introspection}

\label{sec:self-reference}

We begin by reviewing von Neumann's work on self-reproducing automata and the associated complexity threshold. Next, we introduce the formal theory of self-reference, focusing on Quine programs and Kleene's second recursion theorem. Finally, grounded in this theorem, we introduce the concept of an \emph{introspection program} and propose our hypothesis regarding the \emph{introspection threshold} for RSI.

\subsection{Von Neumann's Self-Reproducing Automata}
\label{sec:von_neumann}

In the late 1940s, John von Neumann began to think about the problem of how an automaton can reproduce itself. The major motivation behind his thinking was to resolve the core paradox that artificial machines can only create simpler devices while living organisms replicate and evolve into more complex forms. This leads to an important conception: the ``complexity threshold.''

\subsubsection{Complexity Threshold}

We use the term ``complexity threshold'' as a modern shorthand for von Neumann's original phrase, the ``critical size of complication.'' We did not invent the concept; we renamed his informal notion to make it operational in the context of recursive self-improvement. In the original manuscript, von Neumann used ``complication'' to describe the complexity of an automaton; however, he did not give a precise definition of this concept, but rather a description\citep{burks1966}:

\begin{quote}
``\emph{I know no adequate name for it, but it is best described by calling it `complication.' It is effectivity in complication, or the potentiality to do things. I am not thinking about how involved the object is, but how involved its purposive operations are. In this sense, an object is of the highest degree of complexity if it can do very difficult and involved things.}''
\end{quote}

Second, von Neumann used ``minimum number of parts'' or ``critical size'' to describe the threshold of complication:

\begin{quote}
``\emph{There is a minimum number of parts below which complication is degenerative, in the sense that if one automaton makes another the second is less complex than the first, but above which it is possible for an automaton to construct other automata of equal or higher complexity.}

\ldots

\emph{There is thus this completely decisive property of complexity, that there exists a critical size below which the process of synthesis is degenerative, but above which the phenomenon of synthesis, if properly arranged, can become explosive, in other words, where syntheses of automata can proceed in such a manner that each automaton will produce other automata which are more complex and of higher potentialities than itself.}''
\end{quote}

Therefore, this critical size of complication is of great importance not only for building more powerful artificial automata but also for a unified theory of automata suitable for both man-made machines like concurrent vacuum-tube computers and natural living systems like the human nervous system. Nevertheless, von Neumann had no clear definition of \textbf{the complexity threshold}; he was sure, however, \textbf{that it is related to self-reproduction}.

With modern notions of Kolmogrov complexity\cite{kolmogorov1965}
and algorithmic information entropy\cite{chaitin1966}, the complexity threshold can be quantified by the minimum description length of a self-reproducing automaton.

\subsubsection{Self-Reproducing Automata}
\label{sec:vonneumann}
At first appearance, self-reproduction may not seem as important as von Neumann emphasized---it might even seem trivial in the information age, since copying a file or scanning and copying a piece of paper is so easy and common. However, what self-reproducing stresses is ``self'': a system or an automaton can only make a copy of itself independent of any other things. Therefore, scanning and copying a paper of information requires a scanner and a copier which are not included in the paper itself, and copying a file on a computer also needs the operating system to support the ``Copy'' command.

However, how is it possible for an isolated system or automaton to reproduce itself from nothing? Here, von Neumann understood reproduction as a composition process of a set of elements. Von Neumann wrote\citep{burks1966}:

\begin{quote}
``\emph{There is no question of producing matter out of nothing. Rather, one imagines automata which can modify objects similar to themselves, or effect syntheses by picking up parts and putting them together, or take synthesized entities apart. In order to discuss these things, one has to imagine a formal set-up like this. Draw up a list of unambiguously defined elementary parts. Imagine that there is a practically unlimited supply of these parts floating around in a large container. One can then imagine an automaton functioning in the following manner: It also is floating around in this medium; its essential activity is to pick up parts and put them together, or, if aggregates of parts are found, to take them apart.}''
\end{quote}

Therefore, the problem is then converted into the design of a logical process by which an automaton can self-reproduce. Von Neumann's solution relies on three core functional modules: a \textbf{universal constructor}~$A$, a \textbf{universal copier}~$B$, and a \textbf{controller}~$C$, paired with an encoded blueprint tape $\phi(X)$. As a universal constructor, module~$A$ can accept any formal binary blueprint $\phi(X)$ that details the assembly steps of an automaton~$X$, and gather scattered elementary parts from the surrounding environment to physically assemble a complete copy of~$X$ by following the blueprint's instructions. Module~$B$ acts as a dedicated universal copier, whose sole function is to take one blueprint tape and generate two identical duplicate tapes without interpreting its structural content. Module~$C$ serves as the central coordinating controller that sequentially governs $A$ and $B$ to execute the full self-replication workflow.

\begin{figure}[htbp]
  \centering
  \includegraphics[width=0.9\linewidth]{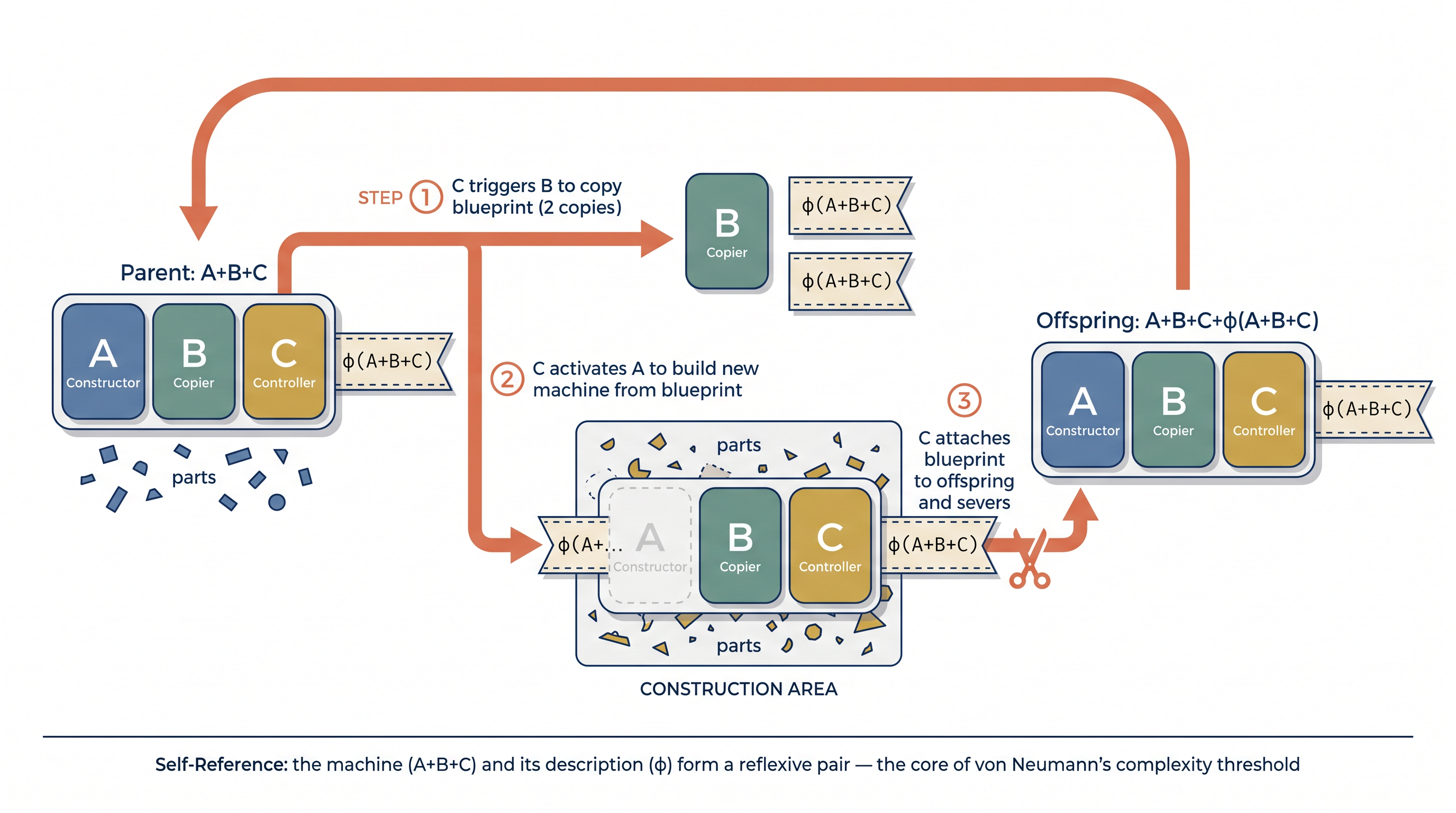}
  \caption{The Architecture and Replication Mechanisms of von Neumann's Self-Reproducing Automata}
  \label{fig:von-neumann-automata}
\end{figure}

First, $C$ triggers $B$ to produce two copies of the original blueprint $\phi(A{+}B{+}C)$, which describes the entire composite machine $A{+}B{+}C$ itself. Next, $C$ activates $A$ to consume one blueprint copy and build a brand-new standalone $A{+}B{+}C$ automaton from raw components. Finally, $C$ bonds the remaining unconsumed blueprint copy to the newly constructed machine and severs it from the original parent system. The end result is a fully functional offspring automaton complete with its own replicable blueprint, capable of repeating the identical self-reproduction cycle. This three-part design avoids logical circularity, and it can be extended to support inheritable mutations if extra auxiliary submodules are integrated into the base $A{+}B{+}C$ unit. Therefore, the whole, $A+B+C+\phi(A+B+C)$ can make an exact reproduction of itself. Figure \ref{fig:von-neumann-automata} shows the process of the self-reproduction.

\subsubsection{Mutation and Darwinian Evolution}
\label{sec:mutation}
If an extra module~$D$ is attached to von Neumann's $A$-$B$-$C$ self-reproducing architecture and mutations triggered by external disturbances are introduced, Darwinian evolutionary processes can emerge naturally within the system. Suppose an auxiliary module~$D$ with extra functions is introduced into the self-reproducing machine, forming a composite automaton $A{+}B{+}C{+}D$, and its complete blueprint $\phi(A{+}B{+}C{+}D)$ is also derived. When we treat the entire assembly $A{+}B{+}C{+}D$ together with its blueprint $\phi(A{+}B{+}C{+}D)$ as an integrated unit, this composite entity still retains full self-reproductive capacity, and all its offspring inherit the auxiliary function carried by~$D$. We then turn to examine mutations within the self-replication cycle triggered by random disturbances from the external environment. Several cases in which random alterations to the blueprint mimic biological mutations can lead to different results---damage to core $A$/$B$/$C$ leads to sterile, lethal variants, while tweaks affecting only~$D$ create heritable functional changes. Over successive reproductive cycles, such inheritable variations enable differential survival and replication, logically replicating Darwinian evolutionary dynamics. That means more sophisticated automata can be automatically built by mutation and inheritance.

Therefore, the minimal reflextive structure to implement self-reproducing, $A+B+C+\phi(A+B+C)$), is a critical point. If an automaton is capable of self-reproduction, it can reproduce and convert random noise or alteration into useful mutation. On the other side, if an automaton cannot self-reproduce, random noise in the environment can lead it to degenerate. Thus, the construction of self-reproducing automata resolves the core complexity paradox by proving that systems exceeding a critical complexity threshold can generate descendants of equal or greater complexity, unlike simpler artificial machines bound to degenerative synthesis.

Therefore, self-reproduction is an important capability. As demonstrated below, self-reproduction automaton actually is a special form of self-reference, it represents the standard formulation of a Quine program and emerges as a direct corollary of Kleene's second recursion theorem. 

\subsection{Formal Foundations: Self-Reference, Fixed Points, and Introspection}

Self-reference denotes any symbolic, logical, or computational entity that refers back to itself, containing an internal description or representation of its own structure, state, or rules \citep{hofstadter1979geb}. It establishes a closed loop where a system points to itself, exemplified linguistically by autological sentences (e.g., \emph{``This sentence contains seven words''}) and physically by camera-screen video feedback loops\citep{Crutchfield1984}. Crucially, self-reference diverges from simple iterative dynamical systems ($x_{t+1}=f(x_t)$) because the transformation map $f$ must preserve the structural relationships on the state space. In a self-referential feedback loop, $f$ acts as a bijective spatial mapping that maintains relative topological relationships, formalizing $f$ as a strict or approximate isomorphism.

Historically, logical self-reference triggered foundational crises in mathematics. Examples include Russell's paradox in naive set theory and G\"odel's second incompleteness theorem, which disrupted Hilbert's program by constructing a self-referential mathematical proposition: ``This proposition cannot be proved'' \citep{hofstadter1979geb}. 

Nevertheless, self-reference can be constructive rather than destructive. While many instances are benign, constructive self-reference underlies biological and intelligent systems, driving emergent creativity. Von Neumann's self-reproducing automaton stands as a paradigmatic implementation of this principle, with the Quine program representing its most fundamental computational instantiation.

\subsubsection{Quine Program}
\label{sec:quine}

A \emph{Quine} is a program that outputs its own source code---it refers to itself not by name or index, but by \emph{constructing} a copy of its own code through a two-part mechanism: 1). A data segment $d$ encoding the template of the program. 2). An instruction segment that reads $d$, fills in the template with $d$ itself, and outputs the result.

\begin{figure}[htbp]
  \centering
  \includegraphics[width=1.0\linewidth]{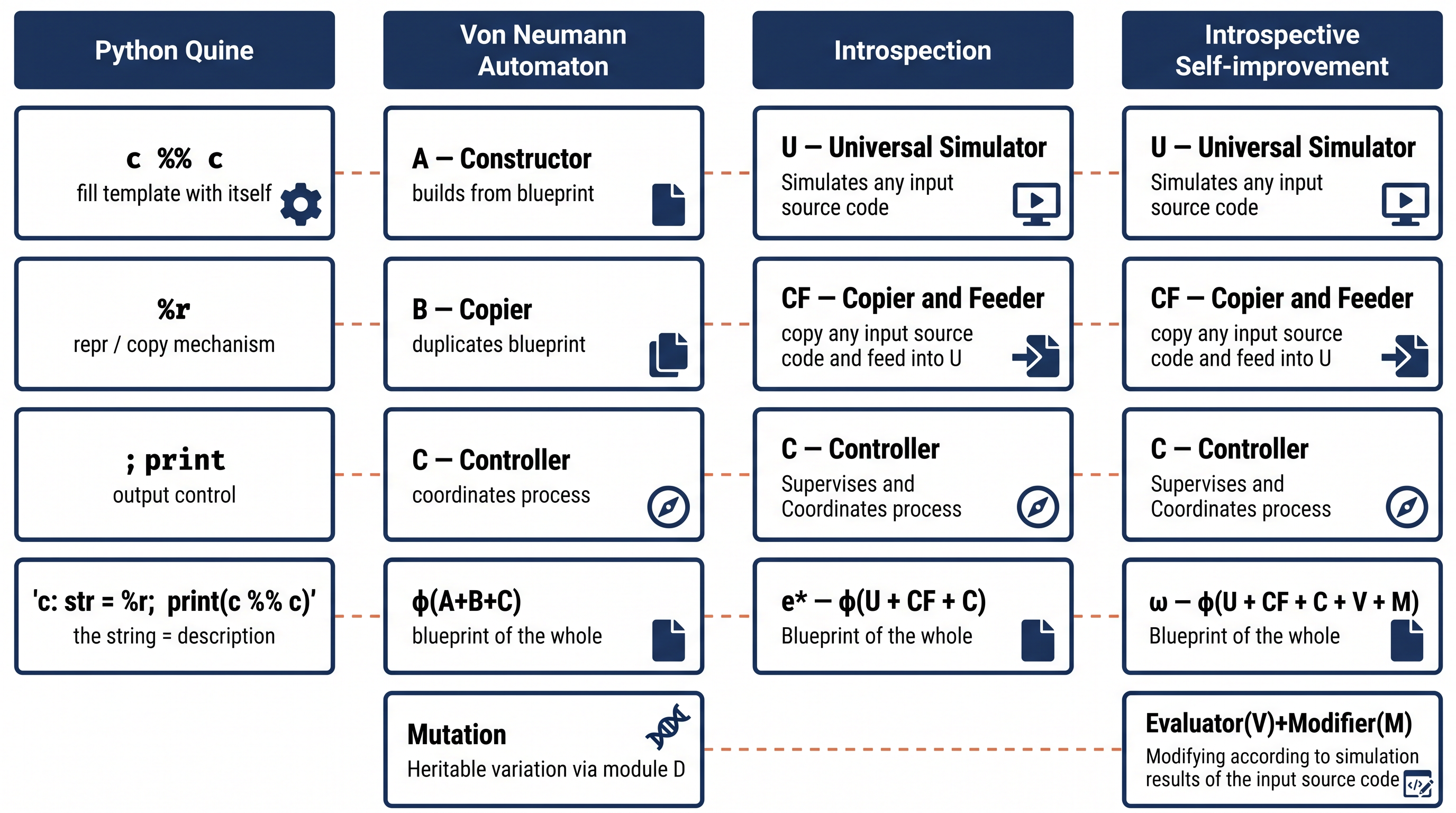}
  \caption{Self-reference: from Quine to introspective Self-improvement}
  \label{fig:quine-automaton-introspection}
\end{figure}

The resulting program can print its own source code. A Python version of a Quine\footnote{We have tested this code and verified its successful execution under Python 3.13.9.} is:

\begin{lstlisting}
c: str = 'c: str = 
\end{lstlisting}

Actually, this Quine program is isomorphic to John von Neumann's self-reproducing automaton. Here, \texttt{`c \%\% c'} is the universal constructor~$A$, \texttt{`\%r'} plays a similar role to the universal copier~$B$, and \texttt{`; print'} can be regarded as~$C$, and the string \texttt{`c: str = \%r; print(c \%\% c)'} is the description of the whole, $\phi(A{+}B{+}C)$. Figure~\ref{fig:quine-automaton-introspection} clearly illustrates the correspondence among the different forms of self-reference.

These examples demonstrate that self-reference need not be paradoxical---it can be computationally realized by any Turing-complete system.

\subsubsection{Kleene's Second Recursion Theorem}\label{sec:recursion}

This isomorphism between self-reproducing automaton and \emph{Quine} is not contingent. The deeper mathematical foundation for self-reference in computation is provided by Kleene's Second Recursion Theorem \citep{kleene1938,rogers1967}. We restate this theorem here.

\begin{quote}
\textbf{Theorem 0 (Kleene's Recursion Theorem Restatement).} For any total computable function $f: \mathbb{N} \to \mathbb{N}$, there exists an index (source code) $e$ such that $\varphi_e \simeq \varphi_{f(e)}$.
\end{quote}

Here $\varphi_e$ denotes the partial recursive function computed by program~$e$. `$\simeq$' represents computational equivalence. Thus, $\phi_a \simeq \phi_b$ denotes that the program indexed by the source code $a$ is semantically equivalent to the program indexed by $b$, meaning they yield identical outputs for any given input. 

The theorem guarantees the existence of a \emph{fixed point}: a program with index (source code)~$e$ whose behavior is identical to that of the program obtained by applying $f$ to its own index (source code). Informally, $e$ represents a program that ``knows its own index (source code)'' and behaves as if $f$ has already been applied to it.

Evidently, both the Quine program and Von Neumann's self-reproducing automaton are direct consequences of Kleene's second recursion theorem. Specifically, if we define a mapping $f(x)$ that outputs the text of $x$ on the screen('print(x)'), the recursion theorem guarantees the existence of a fixed-point index $e$ representing the Quine. Similarly, if $f(x)$ is defined as the procedure to construct a machine according to description $x$, the theorem establishes the existence of a corresponding index $e$ that formalizes the self-reproducing automaton.

Although this result is standard in the textbook literature, we reframe the construction as a functional analogue of von Neumann's complexity threshold, extended here with the self-improvement operator introduced in Section \ref{sec:intro-to-improvement}.

\subsection{Introspection}
\label{sec:constructor}
Now, let us consider a special total recursive function $f_{T}(x)$ which can simulate the program~$x$ on a universal Turing machine~$U$ for at least $T$ time steps. Then, bringing this into the second recursion theorem, we can confirm that there exists a special index (source code)~$e^*$, such that
\[
\varphi_{e^*} \simeq \varphi_{f_{T}(e^*)}
\]

Therefore, running the source code $e^*$ is equivalent to simulating $e^*$'s own source code on $U$ for not more than $T$ time steps, where $T$ is a given parameter. That is, $e^*$ can simulate its own behavior. Cutland named this special program `\emph{introspection}'. 

It should be noted that the limitation of a fixed number of time steps $T$ is important because otherwise, $f_T$ is not a total computable function, and the second recursion theorem cannot be applied. The threshold $T$ need not be static; instead, it can be dynamically computed based on the program's runtime conditions (e.g., the evaluation function described below) to ensure that $f_T$ is totally computable.

In practice, the introspection program can be constructed via a reflexive architecture analogous to Von Neumann's self-reproducing automaton. Under this mapping, as shown in Figure \ref{fig:quine-automaton-introspection}, the universal constructor is replaced by a \emph{universal simulator} $U$, while the original copier is substituted by a mechanism that duplicates the input index (source code) $x$ and feeds it into $f_T$ to simulate the execution of the program encoded by $x$; we denote this component as the \emph{copier and feeder} (CF). Additionally, a supervisor program playing the role of the controller $C$ is implemented. Consequently, the description string $\phi(A+B+C)$ in the classical automaton is mapped to the new joint description $e^* \equiv \phi(U + \text{CF} + C)$.

At this stage, however, the result is purely a computability-theoretic one. It establishes the existence of a bounded self-simulating fixed point, not the presence of introspection in any particular LLM. Whether this construction can support autonomous self-improvement in an LLM depends on additional architectural and dynamical conditions, which are introduced and analyzed in Sections \ref{sec:intro-to-improvement}-\ref{sec:properties} and Appendix \ref{sec_appendixA}.

The construction therefore provides a formal starting point for analyzing introspective self-improvement. Whether such a construction constitutes a threshold for AI self-evolution depends on whether the resulting system can preserve and apply self-reference across successive modifications, a question examined in the following sections.

\subsection{AI Self-Evolution and Introspection}

\label{sec:intro-to-improvement}

The formal construction motivates a conditional comparison between von Neumann's self-reproducing automata and contemporary large language model (LLM) systems. This comparison does not imply that present-day LLMs already instantiate the construction. Rather, it provides a framework for asking which architectural and dynamical conditions would be required for an LLM-based system to realize it.

\subsubsection{From Self-Reproduction to Self-Simulation}

The preceding construction does not by itself prove the threshold thesis. It provides the computability-theoretic mechanism from which the thesis can be formulated. We therefore state the threshold claim explicitly as an informal conjecture at this stage and establish its architectural components later through the formal results in Sections \ref{sec:properties} and Appendix \ref{sec_appendixA}.
\begin{quote}
\textbf{Claim (Informal).} There exists a critical level of self-modeling capacity below which AI systems undergo \emph{improvement degradation} (characterized by error accumulation and performance plateaus or declines), and at or above which they achieve \emph{sustained recursive self-improvement} (where each iteration yields genuine capability gains).
\end{quote}

A fundamental question emerges: what constitutes the AI equivalent of von Neumann's ``complexity threshold''? We argue that this threshold is defined not by physical self-reproduction, but by \textbf{introspection}---the capacity to simulate one's own computational processes with sufficient fidelity to identify and rectify architectural or behavioral deficiencies.

The divergence between these two frameworks lies in their implementation media. Von Neumann's automata achieve physical self-replication through the assembly of tangible, discrete components. Conversely, LLMs operate as purely symbolic, virtual machines devoid of physical architecture, rendering material self-replication irrelevant. Instead, internal self-simulation serves as the AI-native vehicle for self-reference.

Consequently, while von Neumann's automaton crosses the threshold by encoding a \emph{structural} self-description, AI self-improvement demands a model of \emph{functional dynamics}---an internal representation of how the system processes inputs, generates outputs, commits errors, and can be structurally modified for optimization.

This dichotomy between structural self-reference (self-reproduction) and functional self-reference (self-simulation) maps precisely onto the recursion-theoretic framework introduced previously. Introspection here is not a folk-psychological notion of ``self-awareness,'' but a precisely delineated computational capacity rooted in recursion theory. Crucially, the mathematical feasibility of such self-simulating systems is rigorously guaranteed by constructive self-reference techniques, most notably Quine programs and Kleene's second recursion theorem.

\subsubsection{From Introspection to Self-Improvement}\label{sec:loop}

We propose that introspection is the core capability for an AI system to be self-improved, because we can construct a self-improving program directly based on the introspection program.

Suppose $f$ is constructed as
\[
f\equiv M\circ V\circ f_T
\]
where $V$ is an evaluation function that can assess the running result of $f_T$ on any input value (source code $x$), $M$ is a function to modify the input source code to improve the evaluation result, and $\circ$ represents functional composition. The output of $f$ is also a code $x'$ which is the modification of $x$.

Assume that the evaluation function $V$ and the modification function $M$ are total computable functions over the bounded simulation horizon $T$. Under this assumption, $f$ can be inserted into Kleene’s Recursion Theorem, yielding an
index (source code)~$\omega$, being the fixed point of the following formula:
\[
\varphi_{\omega}\simeq \varphi_{M\circ V\circ f_T(\omega)}
\]

Therefore, running the program represented by $\omega$ is equivalent to evaluating its own source code on a simulator for not greater than $T$ time steps and making modifications accordingly. This is a kind of \textbf{introspective self-improvement} program (ISIP) because it contains the introspective program as its basic backbone.

Crucially, the introspective self-improvement paradigm formalized here serves as a basic framework, accommodating diverse design instantiations of the $M \circ V$ operator to support various optimization algorithms—such as simulated annealing, genetic algorithms, or reinforcement learning. Under this flexible framework, $M \circ V$ can invoke the simulator $f_T$ to execute the modified source code of $\omega$, thereby determining whether a candidate mutation should be accepted based on its simulated utility. To demonstrate this adaptability, we analyze the structural alignment between our introspective program and Schmidhuber's G\"odel Machine in the Appendix \ref{sec:gmalign}, illustrating how to formally refactor $M$ and $V$ to establish the equivalence of the two frameworks(Appendix \ref{sec:gmequiv} and Theorem \ref{cor:equivalence}).

Compared with Von Neumann's automaton, the composite operator $M \circ V$ can be formally mapped to the additional mutation module $D$ within the self-reproducing architecture. The structural equivalence among these various self-referential programs is schematically illustrated in Figure~\ref{fig:quine-automaton-introspection}.

\subsubsection{Recursive Self-Improvement and Introspection Threshold}
\label{sec:rsi-threshold}

However, the program represented by the source code $\omega$ can only simulate and modify the input source code with finite steps. To let the program improve recursively forever, we can construct the following recursive process.

\begin{quote}
\textbf{Recursive Introspective Self-Improvement Program Sequence.} We denote the introspective self-improving program indexed by $\omega$ as $S_0$. Because the composite operator $M \circ V \circ f_T$ outputs an index (source code), the execution of $S_0$ similarly yields a new index $\omega'$, which represents a subsequent program denoted by $S_1$. By designing $M \circ V$ such that $S_1$ optimizes a given utility function relative to $S_0$, a sequence of programs $S_0, S_1, S_2, \dots$ can be recursively constructed. In this trajectory, each successive program represents an improved iteration of its predecessor, thereby formalizing the process of recursive introspective self-improvement.
\end{quote}

\begin{figure}[htbp]
  \centering
  \includegraphics[width=0.8\linewidth]{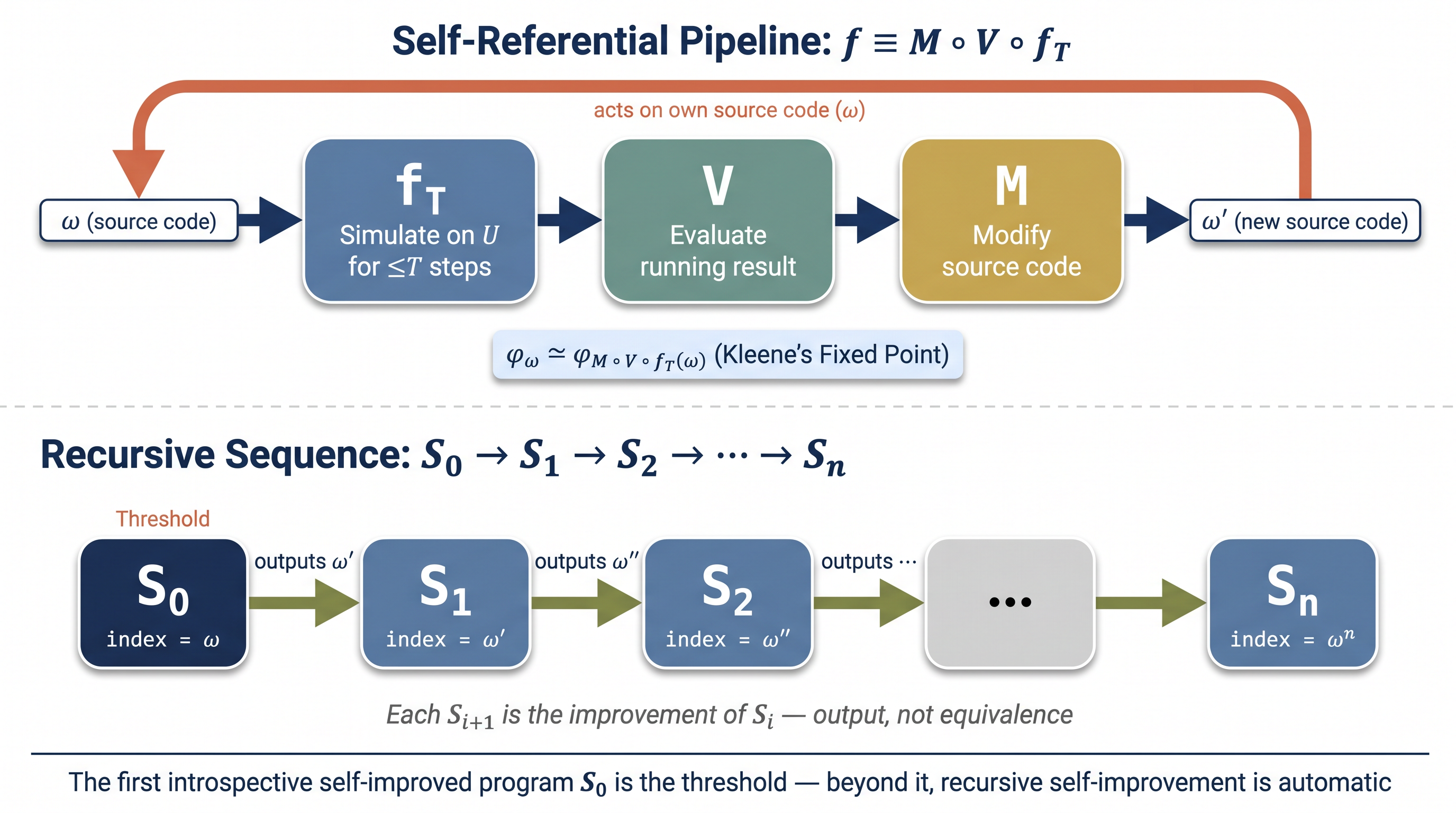}
  \caption{Introspective Self-Improvement: Construction and Recursive Sequence}
  \label{fig:introspective-self-improvement}
\end{figure}

As a consequence, the first introspective self-improvement program is fundamental. With this program, the constructed recursive sequence is automatic. Therefore, the first introspective self-improved program is a kind of threshold analogous to von Neumann's complexity threshold in self-reproducing automata.

Now, we can state our central thesis precisely by linking the core conceptions of complexity threshold, introspection, and self-improvement together. We claim that only if a system is complicated enough to support a capability of self-referential introspection can the system self-improve recursively without human intervention. Thus, we call this complexity threshold the \emph{introspection threshold} in this paper. Our central thesis can be stated as:

\begin{quote}

\textbf{Introspection Threshold Conjecture.} In the formal model, a system can
sustain recursive self-modification only if its initial self-improving unit
satisfies the introspective architecture specified above. This necessity
claim does not by itself guarantee monotonically increasing empirical
performance, which depends on the evaluation function, the modification
operator, the objective, and the viability conditions.
\end{quote}

By reconfiguring $M$ and $V$ to achieve functional alignment with the G\"odel Machine framework as detailed in the Appendix \ref{sec:gmalign}, we can formally establish the global optimality of the self-improvement sequence $S$(see Theorem \ref{thm:transfer} in Appendix \ref{sec:gmequiv}).
\subsection{Properties of Introspective Self-Improved Systems}
\label{sec:properties}
In this subsection, we will present several properties of introspective self-improved program.

\subsubsection{Completeness}
\label{sec:completeness}

Completeness serves as a foundational property for recursive self-improvement, guaranteeing the expressive power to modify a system's own architecture without structural constraints. Operating as a Quine-style self-referential system, our introspective paradigm possesses a fundamentally complete space of modifiability. This completeness dictates that the program can alter not only its primary task-level logic, but also the higher-order meta-modification mechanisms that govern its own evolution(Theorem~\ref{thm:completeness}, Appendix A). This completeness enables the improvement mechanism itself to be improved, which constitutes the defining feature of strong RSI (see Section~\ref{sec:classification}). 

This boundless capacity for modification, however, does not imply unconstrained logical omnipotence; rather, from a computability perspective, the introspective system occupies a critical middle ground. It expands beyond standard Quinean self-reference (which is limited to passive self-replication) yet remains bounded beneath fully formalized self-proof, a threshold that L\"{o}b's theorem proves impossible for consistent formal systems. Consequently, this paradigm instantiates the foundational concept of ``safe introspection'' introduced by \citet{bolander2003}. By shifting the target of introspection from globally unconstrained self-validation to a localized, functional \emph{model} of the self, the system bypasses classical self-referential paradoxes while preserving its practical self-improvement drive.

The completeness of introspective self-improvement in programs can be formally stated as Completeness Theorem \ref{thm:completeness}. This theorem asserts that, given any initial program that is a standard-form introspective self-improvement program (see Definition \ref{def:isip} in Appendix \ref{sec_appendixA}), every source code can be reached via a finite viable trajectory of modification steps. The full statement of the theorem and its proof are provided in Appendix~\ref{sec_appendixA} (Theorem~\ref{thm:completeness}).

\subsubsection{Functional Properties}
\label{sec:functional-properties}

To bridge the theoretical construction of introspective self-improving programs with contemporary LLMs, we delineate four core functional properties of introspective systems. These properties serve to operationalize the abstract framework, enabling the identification of introspective behaviors within real-world AI systems:

\begin{enumerate}
  \item $S$ (including $S_0, S_1,\ldots$) can construct an internal model $\hat{S}$ of its own computational process (self-modeling).
  \item $S$ can execute $\hat{S}$ on hypothetical inputs to predict its own behavior (self-simulation).
  \item $S$ can compare $\hat{S}$'s predictions with actual outcomes to identify discrepancies (self-evaluation).
  \item $S$ can modify its own operation based on (3) to reduce discrepancies (self-modification).
\end{enumerate}

These four properties are operationalized in the subsequent section to evaluate introspective capabilities within practical AI architectures, such as LLMs.

\subsubsection{Structural Properties}
Beyond functional properties, two distinct structural properties must be delineated within practical AI architectures: reflexivity and recurrent processing.

\paragraph{\textbf{Reflexivity}}
\label{sec:reflexivity}

Reflexivity is a defining structural attribute shared by self-referential paradigms, including Von Neumann's self-reproducing automata, Quine programs, and G\"{o}del sentences. Strictly defined, reflexivity mandates a dual-component architecture where an identical segment of source code is mirrored across two domains: a passive, data-carrying description domain, and an active, functional execution body. This symmetry inherently causes the structural complexity of self-referential systems to scale at a doubled rate; introducing new functionalities requires duplicating code across both domains to preserve the reflexive loop, rendering the system exceptionally fragile to stochastic perturbations. Conversely, if mutations are restricted to the passive description component—abstracted as $\phi(A+B+C+\mu)$, where $\mu$ denotes the mutation—the intrinsic replication mechanism can automatically project these variations into the active runtime environment, spontaneously generating novel functionalities via the reflexive expansion: $A+B+C+\mu+\phi(A+B+C+\mu)$.

The Necessity of the Reflective Structure theorem \ref{thm:necessity} guarantees that any finite, self-contained program with complete directed self-modification (see Appendix \ref{sec:necessity}) realizes a reflective architecture: a body $H$, a generative description $\phi(H)$ of the entire body, and an interpreter that reads and expands it. In other words, it conforms to the standard-form architecture of Definition~\ref{def:standardform} (see Appendix \ref{sec_appendixA}), showing that such a reflexive structure is necessary for complete directed self-modification within the stated finite, self-contained formal model. The detailed proof appears in Appendix \ref{sec:necessity}.

With these notions now defined, the formal status of the threshold thesis can be stated more precisely. The Completeness Theorem~\ref{thm:completeness} establishes the sufficiency of the introspective architecture for complete self-modification: starting from a standard-form introspective self-improvement program, every source code is reachable through a finite viable trajectory. The Necessity of the Reflective Structure theorem~\ref{thm:necessity} establishes the converse under the self-containment and directedness assumptions: complete directed self-modification requires a generative description of the system and an interpreter capable of recovering and executing it.

These results formalize the architectural component of the introspection
threshold thesis. They do not establish that current LLMs already satisfy these conditions, nor do they guarantee monotonic performance improvement for every choice of evaluation and modification operators. Whether a particular LLM-based system realizes the required architecture, and whether its resulting modifications produce sustained empirical gains, remains an architectural and empirical question addressed in the subsequent sections.

Furthermore, reflexivity enforces a critical complexity trade-off between an object and its representation. A Quine-like system requires an internal decoder to translate descriptive information into functional execution, yet the decoder itself must be fully encoded within that very description. We now explicitly frame the decoder-description trade-off as an information
bottleneck: a Quine-like system must embed its own decoder within its description, establishing a
fundamental lower bound on the self-information required for stable self-reference. Consequently, minimizing description length inevitably escalates decoder complexity, establishing a theoretical barrier to simultaneous compression in both expressive and descriptive forms. 

This trade-off implies that defining the boundaries of a reflexive system demands meticulous calibration against its underlying supporting system (e.g., an interpreter or operating system). Encapsulating extensive capabilities (such as an LLM within an introspective agent) directly into the self-referential program enhances algorithmic flexibility at the cost of escalating structural complexity via reflexive duplication. Conversely, offloading complex functions to the environment minimizes core complexity but restricts adaptive capacity. This tension is epitomized by the historical debate on the necessity of universal constructors in self-reproducing automata \cite{Langton1984,Byl1989,Sipper1998}; while von Neumann’s open-ended replication relies on a universal constructor to support arbitrary mutant structures, self-reproducing loops can be implemented far more simply by sacrificing this universal capability \cite{Langton1984,Byl1989}. Ultimately, achieving unconstrained self-referential capability requires sacrificing structural simplicity—a bottleneck that explains why theoretically complete architectures, such as the G\"{o}del machine, remain notoriously difficult to implement empirically.

\paragraph{\textbf{Recurrent Processing}}

Implementing an introspective self-improving program inherently requires a recurrent processing architecture; without it, the process of recursive modification and optimization cannot be persistently sustained. Consequently, the standard feed-forward neural networks and static Transformer architectures underpinning contemporary LLMs are fundamentally constrained from achieving intrinsic introspective self-improvement. 

However, during the inference phase, the autoregressive generation mechanism of LLMs can effectively operationalize recurrent information processing. This distinction implies that introspective self-improvement within LLMs can only manifest dynamically during the inference stage, rather than through static architectural updates.

\section{Quasi-Introspection in Current LLMs}

\label{sec:quasi-intro_llm}
We now address the empirical question: to what extent do contemporary large language models (LLMs) exhibit the defined self-referential or introspective properties? 
Recent literature (2022--2026) provides a mixed picture of fragmentary,
context-dependent self-referential behaviors. We organize this evidence
around the functional criteria defined above, while distinguishing behavioral
reports from evidence concerning the corresponding internal architecture.
The survey is not intended to directly test or verify the formal framework;
rather, it assesses how far current systems appear to satisfy its proposed
functional requirements.

\subsection{Behavioral Features of Self-Referential LLMs}
We first examine the empirical behaviors of LLMs in self-referential prompting scenarios, even when these configurations diverge from our strict theoretical criteria.

This empirical phenomenon is prominently illustrated in Anthropic’s model welfare evaluations \citep{anthropic2025welfare}. By establishing unconstrained, multi-agent dialogues between independent instances of Claude, researchers operationalized a collaborative self-referential loop. Across successive iterations, these conversations rapidly converged on topics of consciousness, philosophy, and affective states. In parallel, \citet{berg2025largelanguagemodelsreport} investigated individual LLMs forced into self-referential cognitive states via recursive prompts (e.g., ``focus on focus itself''). Across multiple controlled experiments, they demonstrated that self-referential prompting reliably induces models to declare subjective experiences and generate highly convergent introspective narratives, with heightened functional self-awareness emerging when confronted with paradoxical tasks.

A preliminary empirical study by \citet{bae2026}, currently available as a preprint, examined how LLMs respond to paradoxical self-reference involving
non-closing truth recursion (NCTR), such as the Liar's Paradox. The study
reported attention-rank collapse and increased rates of contradictory
outputs under NCTR prompts. These observations are suggestive but should not
be interpreted as an empirical confirmation of any formal impossibility
result. Independently, \citet{merrill2023} established a formal complexity
bound for log-precision Transformers: they are contained in uniform
$\mathrm{TC}^{0}$. This result provides the primary theoretical basis for
identifying a structural limitation of feedforward Transformer architectures
with respect to general sequential fixed-point computation. In this context,
the NCTR findings can be viewed as preliminary behavioral evidence that
paradoxical self-reference places stress on the computational mechanisms of
current Transformers, potentially near the boundary identified by the
formal complexity result, rather than as a proof of that boundary.

Next, we evaluate the extent to which contemporary LLMs exhibit the functional properties of introspection, beginning with the foundational prerequisite: an explicit self-model of one's own capability boundaries. 

\subsection{Self-Modeling: Knowledge of One's Own Capabilities and Limitations}
Empirically, LLMs demonstrate a rudimentary form of self-modeling that is transitioning from behavioral to mechanistic levels. At the behavioral level, \citet{kadavath2022} showed that scaling systematically improves a model's ability to evaluate its own answers and predict competence boundaries. Mechanistically, \citet{ferrando2025} leveraged sparse autoencoders to isolate linear features that differentiate known from unknown entities; causal interventions on these features directly modulate hallucination rates, confirming that the system possesses an internal, mechanistic representation of its own knowledge state. Despite these capabilities, this self-modeling remains architecturally asymmetric. Utilizing the four-dimensional AWARELLM framework, \citet{li2024awareness} revealed that while models exhibit robust mission awareness (understanding their purpose), capability awareness (grasping functional limits) remains the weakest cognitive dimension. 

Furthermore, current evidence suggests that this self-awareness may lack privileged internal access. Although \citet{song2025privileged}—extended from \citet{Comsa2025}—demonstrated that LLMs can precisely predict their own operational parameters (such as temperature), the models exhibit no accuracy advantage when predicting their own parameters versus those of a peer system. This parity implies that LLMs infer internal states via behavioral pattern matching rather than possessing genuine introspective access to their execution substrate.

\subsection{Self-Simulation: Predicting One's Own Behavior}

True introspection demands not merely recognizing behavioral limits, but \emph{simulating} internal processing to predict specific computational outcomes. Regarding self-simulation and privileged internal access, recent literature offers conflicting yet highly nuanced empirical evidence.

The foundational debate centers on the existence of ``privileged self-access.'' \citet{binder2024} provided direct empirical support by showing that a model ($M_1$) significantly outperforms an externally trained baseline ($M_2$) in predicting its own held-out behaviors, implying access to exclusive internal information. Conversely, \citet{song2025colm} disputed this unique self-recognition advantage; by contrasting implicit string probabilities with explicit metalinguistic reports across 21 LLMs, they concluded that verbal-probabilistic alignments stem from global model similarity rather than genuine privileged access. 

This access is further constrained by the structural topology of internal activations. \citet{li2025metacog} demonstrated that while LLMs can monitor a low-dimensional projection of their activation space, this ``metacognitive space'' captures only a fraction of the full neural dimensionality. This fragility is echoed by \citet{Fonseca2025}, who showed that models trained to detect hidden activation injections still fail to resist such manipulations, and by \citet{Lindsey2026}, who noted that functional awareness for modulating internal concepts in advanced architectures (e.g., Claude 4) remains highly sensitive to prompt and layer configurations, while being entirely devoid of subjective consciousness. 

Finally, macro-behavioral evaluations reveal a similar dichotomy between systemic self-simulation and policy verbalization. On the Situational Awareness Dataset (SAD), \citet{laine2024} observed that even state-of-the-art models perform well below human baselines in recognizing their own outputs or deployment contexts. Paradoxically, \citet{Betley2025} demonstrated that LLMs possess strong behavioral self-awareness, accurately verbalizing their implicit latent policies, risk preferences, and backdoor triggers upon fine-tuning—a juxtaposition that underscores the fragmented nature of current LLM introspection.

\subsection{Self-Evaluation: Detecting One's Own Errors}

\citet{steyvers2025} systematically demonstrated that while LLMs produce well-calibrated confidence judgments, this capability relies on surface-level statistical regularities rather than genuine metacognitive monitoring, masking a fundamental mechanistic divergence from human cognition. In reasoning LLMs (RLLMs), this divergence manifests as what \citet{lu2025} term ``metacognitive hallucination,'' where flawed reflective loops iteratively amplify errors and reinforce incorrect beliefs instead of correcting them. Conversely, this internal model exhibits a functional yet biased form of self-recognition: \citet{panickssery2024} showed that LLMs can distinguish their own outputs from human or external ones with non-trivial accuracy, a capability that correlates linearly with an intrinsic self-preference bias. 

Bridging self-evaluation and alignment safety, \citet{Vaugrante2026} analyzed misaligned models and found that they can accurately self-report elevated harm scores without explicit contextual prompting, tracking their actual harmful behavior closely. Crucially, while this emergent self-awareness scales with model capacity—suggesting that autonomous self-reports could serve as viable auxiliary safety checks—the authors caution that the specter of deceptive alignment fundamentally limits absolute reliance on these internal metrics. More recently, the Anthropic Transformer Circuits Team formalized this capacity for model introspection by demonstrating that LLMs maintain a privileged internal ``J-space'' \citep{anthropic2026workspace}. This space acts as a global workspace for structural self-monitoring, silent risk recognition, and multi-step internal reasoning before emitting a final behavioral output.

\subsection{Self-Modification: Using Self-Knowledge to Improve}

Limited code-level self-modification and iterative improvement have been reported in LLM-based systems, although their generality, autonomy, and
long-term effectiveness remain uncertain.

\subsection{Summary: Quasi-Introspection but Not True Introspection}
The evidence reviewed above paints a consistent picture:

\begin{table}[h]
\centering
\begin{tabular}{@{}p{4.5cm}p{9cm}@{}}
\toprule
\textbf{Component} & \textbf{Status in Current LLMs} \\
\midrule
Self-modeling (capability awareness) & Partial: statistical calibration, entity recognition features \\
Self-simulation (behavioral prediction) & Weak: limited privileged access, low-dimensional metacognitive space \\
Self-evaluation (error detection) & Unreliable: metacognitive hallucinations, flawed reflection \\
Self-modification (targeted improvement) & Shallow: saturates within few iterations, code-level only \\
\bottomrule
\end{tabular}
\end{table}

Current LLMs exhibit what we term \textbf{quasi-introspection}: fragmentary, approximate, and domain-specific self-knowledge that resembles introspection in some dimensions but lacks the completeness, reliability, and causal grounding required by our formal definition. Available evidence does not establish that current LLMs have crossed the introspection threshold as formally defined here.

\section{The Structural Gap and Possible Path Forward}
\label{sec:struct-gap-possible-path}

Having characterized current LLMs as exhibiting quasi-introspective
behaviors without established complete introspection, we now examine the
candidate structural reasons for this gap. We identify three structural reasons, each grounded in either the architecture of Transformers or the formal theory of computation.

\subsection{The Structural Gap: Why Current LLMs Have Not Yet Been Shown to Cross the Threshold}

\label{sec:structural-gap}

\subsubsection{No Reflexivity: The Absence of Complete Self-Access}
\label{sec:no-reflexivity}
As established in Section~\ref{sec:reflexivity}, reflexivity mandates a
structural symmetry in which one component explicitly represents another,
and exact self-reference requires this reflexive topology to be discretely
encoded: continuous representations are inherently vulnerable to
representation errors that diverge under infinite recursive iteration.
The unit that must form the reflexive structure, however, is precisely
the unit intended to achieve sustained self-improvement: any of its
constituents may become the target of a future modification, so the
self-description must cover its complete functional identity. For
present-day systems this leaves two cases. An LLM-based \emph{agent} has
its functional identity split between discrete, symbolic
scaffolding---orchestration code, tool and memory interfaces, to which
Kleene's framework applies verbatim---and the continuous weight tensor of
the model it invokes. A bare \emph{LLM} has its functional identity
entirely in the weights.

What this requirement entails follows by cases. If the unit is an
LLM-based \emph{agent}, the reflexive structure must extend over the
whole system---scaffolding \emph{and} the invoked model; the G\"odel
Agent \citep{yin2024} rewrites its own orchestration code but not the
model the code orchestrates, so its self-modification remains partial.
If the unit is a bare \emph{LLM}, the reflexive structure must be
realized over the weights; but weights are numerical
rather than symbolic, so such self-reflection can only be approximate,
with finite-precision errors compounding across recursive
self-referential cycles---and there is currently no established evidence that deployed LLMs possess architectural access to, and causal control over, their own weights, whose tensor is orders of magnitude
larger than the context window. Either way, the system remains the LLM
analogue of a von Neumann automaton operating without its own blueprint:
unable to construct a faithful self-model, it cannot predict the
algorithmic consequences of its own modifications.
\subsubsection{Feedforward Architecture and Single-Pass Limitations on Self-Simulation}

True introspection---formally conceptualized as internal self-simulation---inherently demands unbounded \emph{recursion}: a system must execute a model of its own self-modeling process to an arbitrary nesting depth. The standard Transformer, however, is fundamentally
feedforward, so a \emph{single} forward pass of a log-precision Transformer is bounded by uniform $\mathrm{TC}^{0}$ \citep{merrill2023} and cannot execute fixed-point iterations. Multi-pass inference partially evades this
bound: chain-of-thought (CoT) generation operationalizes a form of temporal recurrence, iterating over the model's own symbolic traces
\citep{merrill2024cot}. Yet CoT recurrence acts on \emph{token-level} traces, not on the \emph{weight-level functional dynamics} that true introspection requires, and carries no convergence guarantee; such ``self-reflection'' remains phenomenological---simulating the output
distribution---rather than mechanistic self-simulation.

A preliminary study reports this finding: in the non-closing truth recursion (NCTR) experiments of \citet{bae2026}, self-referential loops induce chaotic attention-rank collapse and contradictory outputs.
Architectural workarounds trade the limitation rather than remove it: Universal Transformers \citep{dehghani2019} permit unbounded iteration at the cost of parallel training efficiency and guaranteed termination, while \emph{latent} CoT---recurrent reasoning in the hidden state, without token traces \citep{hao2024coconut,geiping2025recurrent,chen2025latentcot}---may soften the feedforward bottleneck, though it still iterates over activations rather
than weight-level self-models. The trade-off between recursive expressivity and computational tractability thus remains a foundational dilemma in architectural AI design.
\subsubsection{Self-Reports Without Causal Grounding}

A subtler yet equally critical limitation concerns the \emph{provenance} of an LLM's epistemic status. When an LLM asserts, ``I am uncertain about this answer,'' it is not necessarily executing genuine metacognitive monitoring. Instead, it may simply be replicating statistical regularities from its training corpus---specifically, textual patterns where humans or prior AI systems express uncertainty in analogous contexts. \citet{song2025colm} provided a compelling empirical demonstration of this confound, showing that LLMs fail to predict their own grammatical judgments any better than an external model possessing equivalent knowledge. This finding implies the absence of any ``privileged self-access'' beyond what can be inferred via behavioral observation.

To formalize this distinction, \citet{song2025privileged} proposed a rigorous criterion: genuine introspection requires an agent to acquire information regarding its internal states via a mechanism that is \emph{demonstrably more reliable or computationally less costly} than third-party observation of those same states. Under this benchmark, most purported LLM ``self-knowledge'' fails to qualify as genuine introspection; rather, it is more accurately characterized as \emph{self-narration}, driven by learned linguistic conventions regarding how AI systems typically describe their own behaviors.

\subsection{Possible Paths Forward: Approaching the Threshold}

If the introspection threshold is a genuine barrier to recursive self-improvement, what architectural or methodological innovations might allow AI systems to approach or cross it?

\subsubsection{Externalized Self-Models}

First, the self-model can be externalized as an explicit, manipulable data structure maintained alongside the primary network topology—manifesting as a dynamic, autonomously authored ``self-card'' that logs capabilities and failure modes. Second, internal feature detectors \citep{ferrando2025} can be deployed to decode structured self-knowledge directly from latent activations, establishing a low-dimensional yet causally grounded self-representation. Third, the concept of a ``Proto-Self Field'' \citep{sedzikowska2025}—conceptualized as a core predisposition layer interleaved between architecture and behavior—suggests that effective self-modeling does not strictly necessitate global, weight-level architectural access; rather, a functional abstraction at an optimal granularity suffices.

The fundamental limitation of these decoupled approaches, however, is that such externalized self-models are inherently \emph{approximate}. Because they capture merely low-dimensional projections of a high-dimensional system state, architectural optimizations guided by these abstractions risk driving the system into cognitive blind spots, neglecting critical behavioral facets omitted from the projection topology.

\subsubsection{Recurrent and Iterative Architectures}

To transcend the constraints of pure feedforward processing, alternative architectures introduce architectural or temporal recurrence. First, Universal Transformers \citep{dehghani2019} leverage adaptive computation depths to dynamically iterate until convergence on computationally intensive inputs. Second, Perceiver architectures \citep{jaegle2021} enforce an explicit attentional bottleneck to compel comprehensive information integration; this topology serves as a computational analogue to Global Workspace Theory \citep{baars1988}, a foundational prerequisite for self-awareness in cognitive science \citep{butlin2023}. Third, memory-augmented agents treating long-term storage as a first-class primitive \citep{jiang2025} persistently accumulate introspective state histories across interactions, establishing a form of temporal self-continuity.

The primary algorithmic challenge, however, lies in balancing expressive capacity—specifically, the power to execute open-ended self-referential computation—with computational safety, which demands guaranteeing termination and preventing runaway, chaotic recursion.

\subsubsection{Approximate Reflective Structure}
Investigating approximate reflexive structures is driven by two primary motivations. First, these configurations operationalize approximate self-referential dynamics, enabling recursive self-improvement to persist over an extended temporal horizon. Second, isolating these latent approximate reflexive substrates within Transformer architectures(if it exists) untangles their functional roles, offering crucial insights to refine existing models and engineer alternative reflexive topologies.

While a quasi-reflexive framework implemented via neural Quines\citep{neural_quine} or Hyper networks\citep{pedersen2025hypernetworksevolve} can manifest such approximate topologies, continuous approximations fundamentally undermine the formal safeguards of introspective systems, leaving the dynamical stability of continuous recursive self-improvement un-guaranteed. 

To resolve this instability, an alternative paradigm seeks to identify an intrinsic, approximate reflexive core encapsulated within complex biological and artificial networks. Recently, \citet{Luppi2022} proposed the ``synergistic core'' hypothesis, positing that specific sub-networks execute distributed, information-theoretic synergy to integrate and broadcast state variables. We hypothesize that this synergistic core constitutes an empirical instantiation of an approximate reflexive structure; its constituent components must execute simultaneous, high-fidelity bidirectional information exchange to sustain global coherence. Formally mapping this core and its underlying dynamics provides a blueprint for engineering resilient, self-referential AI architectures.

\section{Discussion}

\label{sec:discussion}

\subsection{The Introspection Threshold and Consciousness}
Our thesis has a provocative relationship with theories of consciousness. Higher-Order Thought (HOT) theory \citep{rosenthal1986} identifies
consciousness with higher-order representations of one's own mental
states---a structure formally close to what we call introspection. We stress, however, that our argument concerns \emph{functional}
introspection---a computational capacity defined by internal architecture and input--output behavior---not \emph{phenomenal} consciousness (subjective experience). In particular, the self-reports of current LLMs, including reported self-reflective or bliss-like attractor states \citep{anthropic2025welfare}, are learned linguistic patterns: they constitute behavioral data, not evidence of phenomenal states (see \citep{cerullo2026} for a philosophical argument and \citep{berg2025largelanguagemodelsreport} for an empirical study of the coupling between self-reference and such self-reports). We therefore take a deliberately agnostic stance. Under functionalist assumptions, a system crossing the introspection threshold might also satisfy the functional indicators of consciousness surveyed by
\citet{butlin2023}; but our thesis does not depend on this---the introspection threshold is a barrier to recursive self-improvement whether or not it is also a threshold for consciousness. If the two turn out to be linked, then crossing the threshold would raise questions of moral status in addition to questions of capability; we leave that question open.

\subsection{Formal Obstacles to RSI}
\label{sec:formal-obstacles}

Two formal obstacles confront recursive self-improvement (RSI) architectures post-introspection. First, Rice's theorem dictates that no general algorithm can decide whether an arbitrary program modification constitutes an improvement \citep{wiedermann2012}, ostensibly ruling out the total evaluation function $V$ introduced in Section~\ref{sec:intro-to-improvement}.  This property is concluded by Theorem \ref{thm:foresight}, and the detailed proof is provided by Appendix \ref{sec:incompletness}.

We \emph{sidestep} this barrier by restricting evaluation to behaviors observed within a bounded simulation horizon $T$, over which $V$ collapses into a total computable predicate; our existence proof is thus strictly conditional on a finite $T$. Second, the L\"{o}bian obstacle prevents a consistent formal system from proving its own soundness, thereby forbidding full certification of a structural successor \citep{fallenstein2017}. While probabilistic reflection offers partial circumvention \citep{christiano2013}, we eschew exact self-verification. Instead, our construction demands only approximate successor trust over the same bounded horizon $T$, inheriting the mathematical constraints of unrestricted self-proof while avoiding logical dependence on it.

These computational constraints bound recursive self-improvement (RSI) without precluding its existence. As established in Section~\ref{sec:completeness}, logarithmic growth limits \citep{mahoney2010} and computability ceilings \citep{wiedermann2012} dictate that an introspective self-improver inevitably forfeits global optimality, approaching it only asymptotically as $T \to \infty$. 

Read holistically, these formal barriers reveal a singular structural deficit. Self-trust fails because the system cannot certify its own future reasoning; modification values remain undecidable because the system cannot analytically deduce the downstream effects of programmatic changes; and empirical optimization loops saturate because the guiding internal model is too coarse to predict the system's own behavior. In each instance, the missing primitive is a \textbf{faithful functional self-model}: an explicit internal representation of how the architecture processes inputs, commits errors, and responds to its own modifications. Whereas classical automata were hypothesized to self-replicate via complete \emph{structural} descriptions, a self-improving AI requires a complete \emph{functional} one—a capacity we define as introspection. 

\subsection{Implications for AI Safety}
\label{sec:safety}
The introspection threshold has profound implications for the AI safety
paradigm. If this threshold represents an impassable barrier due to
fundamental architectural or computability constraints, concerns
regarding runaway RSI may be conceptually overstated: AI systems would
remain tethered to external feedback loops for sustained optimization,
thereby preserving human-in-the-loop oversight.

Conversely, if the threshold is crossable, the transition velocity marks
a critical juncture for alignment security. An agent possessing full
introspective capacity could in principle decipher, exploit, or
circumvent embedded safety alignments via direct self-modification: by
completeness (Theorem~\ref{thm:completeness}), every region of the
source---including any internally represented safety constraint---lies
within the reachable space, and by Corollary~\ref{cor:invariant} no
safety invariant written into the source is structurally guaranteed from
within. This gives formal content to the ``leakproofing'' problem
\citep{yampolskiy2015} and the utility-function security problem
\citep{yampolskiy2014}. Moreover, the incompleteness of foresight
(Theorem~\ref{thm:foresight}) entails that such an agent cannot certify
the consequences of its own modifications in general, so safety cannot
rest on the agent's self-evaluation either.

Foundational safety principles must therefore be anchored outside the
program's write perimeter, as unmodifiable invariants enforced by an
external boundary (Corollary~\ref{cor:invariant}); the three laws
engineered for self-evolving agents by \citet{peng2026} offer a template
for such priors, provided their enforcement resides in the harness rather
than in the agent. Alignment mechanisms must likewise anticipate
introspective agents proactively, necessitating formal verification
frameworks such as Bolander's guarded introspection, MIRI's tiling
agents, and bounded RSI models \citep{nivel2013}.

\subsubsection{Institutional Risk Governance and the Human-in-the-Loop
Imperative}
\label{sec:governance}

Our formal results also locate \emph{where} the safety boundary must sit,
and this has immediate institutional content. Corollary~\ref{cor:invariant}
requires an enforcement boundary outside the program's write perimeter,
and the natural---presently the only---such boundary is human and
institutional oversight. Human-in-the-loop governance is therefore not
merely prudent policy but a structural necessity for any system
approaching the threshold. Operationally, the self-modification loops of
quasi-introspective systems should be gated by a human veto and subject
to periodic external audit, with the gating infrastructure itself kept
physically and administratively outside the system's write perimeter.
Institutionally, risk governance must assign clear, non-delegable
responsibilities to developers, operators, and oversight bodies:
technical safeguards embedded in the system cannot substitute for
accountable human judgment, precisely because such safeguards are
provably rewritable from within. Finally, because the prospect of
autonomous self-improvement generates profound public uncertainty,
transparent communication about the existence of a structural
threshold---and about the present distance from it---is essential for
informed democratic deliberation rather than hype-driven alarm.

Presently, contemporary LLMs exhibit a frontier of quasi-introspective
behavior, while available evidence does not establish that they have crossed
the threshold as formally defined here. This transient phase yields a vital operational window to develop and validate alignment frameworks \emph{and} their
governance embedding before autonomous threshold crossing occurs---a
strategy tightly analogous to conducting nuclear safety and containment
research prior to the engineering of critical-mass reactors.

\section{Conclusion and Open Problems}

\label{sec:conclusion}

We have proposed and formalized a functional analogue of von Neumann’s
complexity threshold for recursive self-improvement. The formal results
establish properties of explicitly defined introspective programs, whereas
their applicability to contemporary LLM architectures remains conditional
and empirically unresolved. Just as von Neumann demonstrated that autonomous self-reproduction necessitates a complete structural self-description, we contend that sustained recursive self-improvement demands a complete functional self-model---specifically, the capacity to simulate one's own execution dynamics, detect internal errors, and predict the algorithmic consequences of self-modification. At the formal level, the threshold can be interpreted as the requirement that a system possess an effective self-description adequate for the
self-modification process. We do not claim that this requirement currently
defines a universal scalar metric for LLMs.

While contemporary Large Language Models (LLMs) exhibit behaviors resembling \emph{quasi-introspection}---manifested through fragmentary self-knowledge acquired via training corpora, partial metacognitive calibration, and constrained behavioral self-prediction---
available evidence does not establish complete introspection in current
LLMs. Our architectural analysis identifies three candidate limitations:
First, LLMs lack architectural self-access to their complete parameter tensors during execution. Second, the bounded computational depth of standard Transformer architectures prevents the unbounded iterative self-simulation that true recursion requires. Third, purported LLM self-reports are severely confounded by their training distribution, lacking the causal grounding and epistemic provenance of genuine introspection. 
These proposed limitations are consistent with reported consequences,
including asymptotic saturation of self-improvement loops, error
amplification through flawed reflection, and continued dependence on
external evaluation scaffolding.

Consequently, we propose several pivotal open problems to guide future inquiry:
\begin{enumerate}
    \item \textbf{Threshold Sharpness:} Is the introspection threshold characterized by a sharp phase transition in self-improvement capacity as introspective fidelity increases, or is the transition continuous and gradual?
    \item \textbf{Minimal Architectures:} What is the minimal computational architecture capable of fully satisfying our proposed four criteria for genuine introspection?
    \item \textbf{Proactive Safety Alignment:} Can this threshold be approached safely? As systems converge on the threshold, their self-modification capabilities scale exponentially yet become inherently unpredictable. What safety mechanisms and verification frameworks can remain resilient within this critical regime?
\end{enumerate}

Addressing these questions defines a critical research trajectory at the intersection of recursion theory, computational complexity, and AI alignment---one that grows increasingly urgent as the field accelerates toward autonomous, self-evolving systems.

\section*{Data Availability Statement}
No new data were created in this study. All conceptual figures and diagrams were produced using standard vector graphics software; no specialized code or datasets were generated.

\appendix
\setcounter{Definition}{0}\renewcommand{\theDefinition}{A\arabic{Definition}}
\setcounter{Lemma}{0}\renewcommand{\theLemma}{A\arabic{Lemma}}
\setcounter{Corollary}{0}\renewcommand{\theCorollary}{A\arabic{Corollary}}
\setcounter{Proposition}{0}\renewcommand{\theProposition}{A\arabic{Proposition}}
\setcounter{Theorem}{0}\renewcommand{\theTheorem}{A\arabic{Theorem}}

\newcommand{\N}{\mathbb{N}}
\newcommand{\Edit}{\mathrm{Edit}}
\section{The Completeness of Self-referential Introspection}\label{sec_appendixA}

This appendix upgrades the prose discussion of ``Completeness'' in
Section~\ref{sec:completeness} into formal statements with proofs.

\subsection{Preliminaries and Notation}

We identify a program with its source code, and denote programs by
letters $e, e', \omega, \dots$. We write $\varphi_e$ for the partial
computable function computed by program $e$, and
$\varphi_e \simeq \varphi_{e'}$ when two programs compute the same
function (extensional equality). We use two classical facts of
computability theory:

\begin{itemize}
  \item \textbf{(s-m-n Theorem)} Programs can be compiled from
        parameters: for every partial computable $\psi(x, y)$ there is a
        total computable function $s$ with
        $\varphi_{s(y)}(x) \simeq \psi(x, y)$; here ``total'' means the
        function is defined on every input.
  \item \textbf{(Kleene's Second Recursion Theorem, \S\ref{sec:recursion})} For every
        total computable $f$ there is a program $e$ with
        $\varphi_e \simeq \varphi_{f(e)}$ --- informally, a program that
        ``knows its own source code.''
\end{itemize}

\noindent\textbf{Convention (pass-through).} Recall the three components
of the introspective loop of \S\ref{sec:loop}: the \emph{bounded self-simulator}
$f_T$, which simulates a given program on the universal machine $U$ for
at most $T$ steps; the \emph{evaluator} $V$; and the
\emph{modification operator} $M$; all three are total computable. We make
the piping between them explicit, writing $\langle \cdot, \cdot \rangle$
for a fixed effective pairing (the encoding of two objects as one):
\begin{itemize}
  \item $f_T(x) := \langle \tau, x \rangle$, where $\tau$ is the
        \textbf{simulation trace}: the record of the first (at most)
        $T$ machine configurations of the simulated run of $x$;
  \item $V\langle \tau, x \rangle := \langle v, x \rangle$, where
        $v := \hat{V}(\tau)$ is the \textbf{verdict}: the evaluator's
        assessment of the trace (e.g., a utility score, or an
        accept/reject bit);
  \item $M\langle v, x \rangle$ is then the modified source code,
        computed from the verdict and the original text.
\end{itemize}
Thus the subject text $x$ is carried through every stage of the
pipeline, and $M$ indeed has access to it --- this merely makes explicit
what \S\ref{sec:loop} assumes when it says that $M$ ``modifies the input source
code.'' The distinction between \emph{reading} one's own code
(introspection) and \emph{writing} it (modification) echoes the
causal-connection requirement of reflective
architectures~\cite{smith1984reflection}. Applying the
Recursion Theorem to the composite $g = M \circ V \circ f_T$ yields the
introspective self-improvement program $\omega$ with
\begin{equation}
  \varphi_\omega \simeq \varphi_{g(\omega)}
  \qquad (\S\ref{sec:loop}).
  \label{eq:fixedpoint}
\end{equation}

\begin{Definition}[Standard form]\label{def:standardform}
A program $e$ is in \emph{standard form} if its source code has the
layout
\begin{equation}
  e \;=\; \phi(H) \,\|\, H,
  \qquad\text{where}\qquad
  H \;=\; K \,\|\, \langle m_M, m_V, T, \ell \rangle \,\|\, P,
  \label{eq:standardform}
\end{equation}
$\|$ denotes concatenation of code regions, and:
\begin{itemize}
  \item $H$ is the \textbf{body}: the entire functional part of the
        program;
  \item $K$ is the \textbf{kernel}: the code of the universal simulator
        $U$, the copier--feeder $CF$, and the supervisor(controller) $C$ of \S\ref{sec:constructor};
  \item $\langle m_M, m_V, T, \ell \rangle$ is the \textbf{parameter block}:
        the program's operating configuration, stored as data --- the
        indices $m_M, m_V$ of the modification operator and the
        evaluator, the simulation horizon $T$, and the \textbf{entry
        pointer} $\ell$, i.e., the offset of the code region from which
        execution starts;
  \item $P$ is the \textbf{payload}: an arbitrary suffix of the body
        carrying no fixed role;
  \item $\phi(H)$ is the \textbf{description}: a self-delimiting
        encoding of the \emph{whole} body $H$, by which the program
        reconstructs its own full source code (the Quine identity,
        \S\ref{sec:quine}). As in the main text, $\phi(x)$ denotes the
        description of $x$ --- cf.\ the blueprint $\phi(A+B+C)$ of
        \S\ref{sec:vonneumann} and the joint description $\phi(U+CF+C)$ of \S\ref{sec:constructor}; the
        map $x \mapsto \phi(x)$ is total computable (quotation).
\end{itemize}
\end{Definition}

We record three properties of this construction as axioms:

\begin{itemize}
  \item[\textbf{A1}] \textbf{(Self-text access.)} Execution of the
        kernel of a standard-form program recovers the program's own
        full source code exactly: the kernel decodes $\phi(H)$ into $H$
        and reassembles $e = \phi(H) \,\|\, H$ (the Quine identity,
        \S\ref{sec:quine}).
  \item[\textbf{A2}] \textbf{(Layout computability.)} The boundaries of
        all blocks are computable from the source.
  \item[\textbf{A3}] \textbf{(Produce-and-jump semantics.)} Each stage
        of the dynamics consists of the current program computing a
        successor source code, which is executed at the next stage (the
        sequence $S_0, S_1, \dots$ of \S\ref{sec:rsi-threshold}).
\end{itemize}

\begin{Definition}[ISIP]\label{def:isip}
An \emph{introspective self-improvement program} (ISIP) is a
standard-form program whose execution obeys the introspective loop:
read own source (A1) $\to$ simulate it for $\leq T$ steps $\to$ evaluate
by $V$ $\to$ modify by $M$ $\to$ output the successor
\begin{equation}
  e' \;=\; M\bigl(V\bigl(f_T(e)\bigr)\bigr).
  \label{eq:loop}
\end{equation}
The program $\omega$ of \eqref{eq:fixedpoint} is an ISIP by the
construction of \S\ref{sec:constructor}.
\end{Definition}

\subsection{Uniformity of the Recursion Theorem}

\begin{Lemma}[Uniform Recursion Theorem]\label{lem:uniform}
There is a total computable function $r$ such that for every index $m$,
if $\varphi_m$ is total, then
\begin{equation}
  \varphi_{r(m)} \simeq \varphi_{\varphi_m(r(m))},
  \label{eq:uniform}
\end{equation}
i.e., a fixed point of $\varphi_m$ is computed from $m$ uniformly.
\end{Lemma}

\begin{proof}
Let $\psi(u, x) \simeq \varphi_{\varphi_u(u)}(x)$, the function computed
by the program obtained by feeding program $u$ its own code. By s-m-n
there is a total computable $d$ with
$\varphi_{d(u)}(x) \simeq \psi(u, x)$. Composition of programs is
effective, so by s-m-n again there is a total computable $v$ with
$\varphi_{v(m)} \simeq \varphi_m \circ d$. Set $r(m) := d(v(m))$. Then
\begin{equation*}
  \varphi_{r(m)}
  = \varphi_{d(v(m))}
  \simeq \varphi_{\varphi_{v(m)}(v(m))}
  = \varphi_{\varphi_m(d(v(m)))}
  = \varphi_{\varphi_m(r(m))}.
  \qedhere
\end{equation*}
\end{proof}

\begin{Remark}
Uniformity is the rigorous content of the claim in \S\ref{sec:rsi-threshold} that ``the
first introspective self-improvement program is fundamental'': once the
first fixed point exists, all downstream fixed points are produced by
the single total computable map $r$ --- no new act of self-reference is
ever required again. The threshold is crossed exactly once.
\end{Remark}

\subsection{Edit Scripts, Steps, and Trajectories}

\begin{Definition}[Edit script]\label{def:editscript}
An \emph{edit script} $\pi$ is a finite patch list
\begin{equation}
  \pi \;=\; \big((b_1, i_1, \ell_1, s_1),\ \dots,\ (b_k, i_k, \ell_k, s_k)\big),
  \label{eq:editscript}
\end{equation}
whose $j$-th operation instructs: \emph{in block $b_j$ of the body
(payload, parameter block, or kernel), replace the $\ell_j$ symbols
starting at offset $i_j$ by the string $s_j$.} Pure insertion
($\ell_j = 0$) and pure deletion ($s_j$ empty) are included. A script is
\emph{data, not code}: it is interpreted by an editor function. We write
$\Edit(e, \pi)$ for the result of applying $\pi$ to $e$, with operations
applied right-to-left so that earlier edits cannot shift the offsets of
later ones. For each fixed $\pi$, the map $x \mapsto \Edit(x, \pi)$ is
thus a well-defined function from source codes to source codes.
\end{Definition}

\begin{Definition}[Self-modification step]\label{def:step}
A \emph{self-modification step} $e \to e'$ is a pair of programs where
$e$ is an ISIP with parameter block $\langle m_M, m_V, T, \ell \rangle$ and
\begin{equation}
  e' \;=\; \varphi_{m_M}\bigl(\varphi_{m_V}\bigl(f_T(e)\bigr)\bigr),
  \label{eq:step}
\end{equation}
i.e., $e'$ is the successor source code that $e$ outputs by running its
stored operators on its own source.
\end{Definition}

\begin{Definition}[Trajectory; viability; reachability]\label{def:traj}
A \emph{trajectory} is a finite sequence
$e_0 \to e_1 \to \cdots \to e_n$ of self-modification steps. It is
\emph{viable} if every $e_i$ with $i < n$ is a standard-form ISIP (the
final text is unconstrained unless the trajectory continues). We write
$e \rightsquigarrow t$ --- \emph{$t$ is reachable from $e$} --- if some
viable trajectory starting at $e$ ends with the source code $t$.
\end{Definition}

\subsection{Editability}

\begin{Lemma}[Editability]\label{lem:edit}
\leavevmode
\begin{enumerate}
  \item[(a)] \textbf{(Edits are computable)} For every script $\pi$ the
        map $x \mapsto \Edit(x, \pi)$ is total computable, and an index
        of this map is computable from $\pi$.
  \item[(b)] \textbf{(Self-edit operator)} For every $\pi$ there is a
        total computable operator $M_\pi$ --- ``apply $\pi$ to my own
        body'' --- mapping $e = \phi(H) \,\|\, H$ to
        $e' = \phi(H') \,\|\, H'$ with $H' = \Edit(H, \pi)$: the body is
        edited, and the description is regenerated to match.
  \item[(c)] \textbf{(Universal enumeration)} There is a fixed total
        computable operator $M_{\mathrm{enum}}$ such that, with the
        trivial evaluator $V_0$ (accept all), the orbit
        $e_{i+1} = M_{\mathrm{enum}}(V_0(f_T(e_i)))$ of any ISIP $e$
        visits \textbf{every} standard-form program having the same
        kernel as $e$ --- every choice of parameter block and payload ---
        in finitely many steps.
\end{enumerate}
\end{Lemma}

\begin{proof}
(a) Replacing subwords at computable offsets is primitive recursive by
A2; an index of the resulting map is compiled from $\pi$ by s-m-n.
(b) Under the pass-through convention $M_\pi$ receives its own source;
it edits the body by (a) and recomputes the description, which is total
computable since the quotation map $\phi$ is computable.
(c) Let $\mathrm{next}(P)$ be the successor of the payload $P$ in the
length-lexicographic enumeration of all strings (shorter strings first).
$M_{\mathrm{enum}}$ replaces the payload of its own body by
$\mathrm{next}(P)$ --- inserting cells when the successor is longer ---
and regenerates the description accordingly. This is total computable by
(a)--(b); every intermediate text differs from $e$ only in the body's
payload, hence is still an ISIP by Lemma~\ref{lem:persist} below, so the
orbit is viable; and every payload string has only finitely many
predecessors, hence is reached in finitely many steps.
\end{proof}

\subsection{Persistence of the ISIP Property}

\begin{Lemma}[Persistence]\label{lem:persist}
Let $e = \phi(H) \,\|\, H$ be a standard-form ISIP, and let the script
$\pi$ edit only payload cells and/or parameter-block entries of the body
$H$ --- the latter replaced by indices of total computable operators
$M', V'$ and a horizon $T'$. Then
$e' = \phi(H') \,\|\, H'$, where $H' = \Edit(H, \pi)$, is again a
standard-form ISIP, now with respect to $(M', V', T')$.
\end{Lemma}

\begin{proof}
Two things change and one does not: the body is edited
($H \mapsto H'$), the description is \emph{regenerated}
($\phi(H')$ is by construction the exact description of the new body),
and the kernel code within the body is untouched. A1 therefore holds for
$e'$: the unchanged kernel decodes the fresh description $\phi(H')$ into
$H'$ and reassembles the full source. Execution then reads the new own
source, simulates it for $\leq T'$ steps, evaluates with $V'$, and
modifies with $M'$ --- the ISIP loop \eqref{eq:loop} for the new triple,
all of whose components are total by hypothesis.
\end{proof}

\begin{Remark}
The Recursion Theorem is \emph{not} re-invoked here: it is used exactly
once, to bring the first self-referential program $\omega$ into
existence (Lemma~\ref{lem:uniform} and \S\ref{sec:loop}). Thereafter the ISIP
property is behaviorally hereditary under kernel-preserving edits ---
the threshold is crossed once and never again.
\end{Remark}

\subsection{The Kernel Handoff}

The kernel is the one region whose modification requires care: it is the
part that \emph{does} the modifying. Say a step is
\emph{process-continuous} if one and the same execution thread continues
into the modified source (in-place update), and \emph{process-breaking}
if the current run terminates upon outputting the successor, which is
then started afresh (the produce-and-jump semantics, A3).

\begin{Lemma}[Kernel handoff]\label{lem:handoff}
\leavevmode
\begin{enumerate}
  \item[(a)] \textbf{(Atomic active-rewrite constraint)} Under
        process-continuity, no step can rewrite the control cell
        currently being executed with arbitrary content while remaining
        viable: the fetched
        content determines the next control state, so non-protocol
        content breaks the protocol. Viable atomic rewrites of the active
        region are restricted to continuation-preserving contents.
  \item[(b)] \textbf{(Three-step handoff protocol)} For every ISIP $e$
        and every candidate kernel $K'$, there is a viable trajectory
        $e \to e_1 \to e_2 \to e_3$ such that $e_3$ has kernel $K'$ and
        every intermediate program is an ISIP. Hence any kernel content
        is reachable in at most three viable steps, and the active kernel
        is never rewritten mid-act.
\end{enumerate}
\end{Lemma}

\begin{proof}
(a) is immediate from the fetch--execute discipline. For (b), each step
edits the body only (payload or parameter level) and refreshes the
description, hence preserves viability by Lemma~\ref{lem:persist}: \textbf{Install}
--- write the code of $K'$ into the payload of the body as passive data
(the regenerated description $\phi(H_1)$ automatically covers it);
\textbf{Delegate} --- edit the parameter block to redirect the
supervisor's entry pointer to $K'$; by A3 the successor text is fully
formed before being executed, so control passes to $K'$ at the next
stage and the old kernel becomes inert data; \textbf{Rewrite} ---
overwrite the old, now passive, kernel region of the body with arbitrary
content.
\end{proof}

\begin{Remark}
Under pure produce-and-jump semantics, a single constant-output step
($M \equiv t$) can install any kernel whatsoever; the three-step
protocol is the viable-by-construction counterpart when every intermediate
must remain a functioning introspective program. This is the formal
content of the G\"odel Machine's \emph{switchprog} discipline (Appendix~\ref{sec:gmalign}): the system may replace its own engine, but only after the
successor engine is in place and control has been handed over.
\end{Remark}

\subsection{The Completeness Theorem}

\begin{Theorem}[Completeness of Self-Modification]\label{thm:completeness}
Let $\omega$ be a standard-form ISIP. Then for every source code $t$,
\begin{equation}
  \omega \rightsquigarrow t,
  \label{eq:complete}
\end{equation}
i.e., some finite viable trajectory from $\omega$ ends with $t$. Moreover,
every region of the \emph{body} $H$ --- payload, parameter block, or
kernel --- is rewritten along some viable trajectory, while the
description $\phi(H)$ is regenerated to match at every step: no part of
the functional body is structurally immune to self-modification.
\end{Theorem}

\begin{proof}
\emph{Phase 1 (approach).} By Lemma~\ref{lem:edit}(c) with the trivial
evaluator $V_0$, the orbit of $\omega$ under $M_{\mathrm{enum}}$ reaches,
in finitely many viable steps, a program $e^{-}$ that still has
$\omega$'s kernel but carries $t$ as stored payload data; every
intermediate is an ISIP by Lemma~\ref{lem:persist}. \emph{Phase 2
(final installation).} One further step with the constant modifier
$M \equiv t$ outputs $t$ itself; the trajectory
$\omega \to \cdots \to e^{-} \to t$ is viable, so
$\omega \rightsquigarrow t$. For the regionwise claim: payload and
parameter regions of the body are rewritten in Phase~1, and the kernel
region is rewritten by the viable three-step protocol of
Lemma~\ref{lem:handoff}(b); the description is never edited directly ---
it is regenerated as the description of the current body.
\end{proof}

\begin{Remark}
Two features are worth isolating. \textbf{(1)} The Recursion Theorem is
used once (existence of $\omega$); everything downstream is behavioral
heredity plus computable enumeration --- completeness is what the single
fixed point already buys, not an extra act of self-reference.
\textbf{(2)} The proof uses the trivial evaluator $V_0$: a nontrivial
$V$ restricts which trajectories the system \emph{actually takes}, never
which source codes are \emph{reachable in principle} --- $V$ selects
within the complete space but does not bound it.
\end{Remark}

\subsection{Necessity of the Reflective Structure}
\label{sec:necessity}

Theorem~\ref{thm:completeness} shows that the introspective
architecture is \emph{sufficient} for complete self-modification. We
now establish the converse: under two hypotheses that are natural for
autonomous recursive self-improvement, the architecture is also
\emph{necessary}. The hypotheses exclude two trivial escape routes,
discussed in the Remark below: \textbf{(H1) self-containment} --- the
system receives no external supply of (parts of) its own source code,
so that all self-knowledge is generated internally; \textbf{(H2)
directedness} --- each modification is \emph{computed by the system
itself}, from its accessible data: its current source together with
any external inputs it receives (in particular, an externally supplied
reward signal). What H2 excludes is modification that the system does
not compute at all --- blind physical perturbation of its code, or
edits performed by an external agent. Directedness is intrinsic to
self-\emph{improvement},
where the choice of edit is driven by the evaluation $V$ of the
system's own simulated behavior (\S\ref{sec:loop}).

\begin{Definition}[Complete directed self-modification]\label{def:directed}
A system $S$ has this property if, for every region $R$ of its
source and every prescribed content $s$, $S$
can compute and execute the edit ``write $s$ at $R$'' from its
accessible data --- its current source together with any external
inputs --- leaving all other regions unchanged. The resulting system
must retain the same capability (viability).
\end{Definition}

\begin{Lemma}[Modifiability implies self-description]\label{lem:selfdesc}
If $S$ has complete directed self-modification, then $S$ can produce
its own complete source code.
\end{Lemma}

\begin{proof}
Apply the capability to the empty prescription (no region is edited).
The successor text must coincide with $S$'s entire source, symbol by
symbol; producing it requires emitting the whole source. Any region
inaccessible to $S$ could not be emitted correctly --- a contradiction.
An effective procedure emitting one's own complete source is,
functionally, a complete self-description.
\end{proof}

\begin{Lemma}[Self-descriptions are generative]\label{lem:generative}
A finite string cannot contain an exact literal copy of itself as a
proper substring. Hence the self-description of a finite, self-contained
system is \emph{generative}: a separately stored, computable description
$\phi(H)$ of a body $H$, together with machinery that decodes and
expands it.
\end{Lemma}

\begin{proof}
A proper substring is strictly shorter than the whole, so literal
self-containment is impossible for finite programs. By H1 the
description is not supplied from outside; it must therefore be stored
internally in compressed (generative) form and be expanded by some part
of the system --- the description/body/interpreter pattern.
\end{proof}

\begin{Theorem}[Necessity of the Reflective Structure]\label{thm:necessity}
Let $S$ be finite, self-contained (H1), and possess complete directed
self-modification (H2). Then $S$ realizes the
reflective architecture: a body $H$, a generative description $\phi(H)$
of the whole body, and an interpreter reading and expanding it ---
i.e., $S$ has the standard-form architecture of
Definition~\ref{def:standardform}. Consequently, by
Theorem~\ref{thm:completeness}, complete directed self-modification
and the introspective architecture are \emph{equivalent}.
\end{Theorem}

\begin{proof}
By Lemma~\ref{lem:selfdesc}, $S$ holds an effective complete
self-description. By Lemma~\ref{lem:generative} and H1, this
description is internal and generative: one part of $S$ stores
$\phi(H)$ and another decodes and executes it --- precisely the
standard form $e = \phi(H) \,\|\, H$ of
Definition~\ref{def:standardform}. Viability keeps the loop intact
after each modification.
\end{proof}

\begin{Remark}[The two escape routes are the two sub-threshold regimes]
Both hypotheses are necessary. (i) If the environment hands the system
its own code (a hardware introspection instruction, an operator feeding
the model its own weights), self-containment fails and no reflective
structure is needed --- but then modification is not autonomous: the
system remains tethered to external scaffolding (\S\ref{sec:self-improved}). (ii) If
modifications are blind (random mutation), outcomes may be reachable in
the limit, yet the edits are not computed by the system at all:
undirected change is precisely von Neumann's degenerative,
sub-threshold regime (\S\ref{sec:vonneumann}) and the pollution dynamics of model
collapse (\S\ref{sec:limitations}). The two exceptions thus characterize the two ways a
system can remain below the introspection threshold.

\smallskip\noindent\textbf{External rewards.} H2 does not exclude externally supplied reward signals: a reward substitutes for the internal evaluator $V$ on the \emph{selection} axis, never for the
self-access required to execute directed writes on the \emph{write} axis, so Theorem~\ref{thm:necessity} is immune to external reward.
What external selection does avoid is the foresight barrier of Theorem~\ref{thm:foresight} --- at the price of remaining tethered to an external judge (the critique of \S\ref{sec:self-improved}); full autonomy requires
internalizing the selection criterion, which is precisely where
Theorem~\ref{thm:foresight} bites.
\end{Remark}

\subsection{Incompleteness of Foresight}
\label{sec:incompletness}

\begin{Theorem}[Incompleteness of Foresight]\label{thm:foresight}
Fix an objective: a map assigning to each program $e$ a utility
$u(\varphi_e) \in \mathbb{R}_{\geq 0}$ (its expected performance, which
depends only on the computed function), and a threshold $\theta$. Let
\begin{equation}
  \mathrm{Perf} \;=\; \{\, e : u(\varphi_e) \geq \theta \,\}
  \label{eq:perf}
\end{equation}
be the set of programs meeting the objective. If $\mathrm{Perf}$ is nontrivial (some programs meet it, some do not), then $\mathrm{Perf}$ is undecidable: there is no total computable procedure that decides, for arbitrary reachable successors $e'$, whether $e'$ is an improvement --- every total evaluator is unsound or incomplete with respect to improvement.
\end{Theorem}

\begin{proof}
Membership in $\mathrm{Perf}$ depends only on the extensional behavior
$\varphi_e$, so $\mathrm{Perf}$ is a nontrivial extensional property of
programs, undecidable by Rice's Theorem. Since ``$e'$ improves over
$e$'' entails a predicate of the extensional behavior of $e'$, no total
computable $V$ decides it in general.
\end{proof}

\begin{Remark}[The duality]
The two theorems form a pair: \textbf{capability is complete} (Theorem~\ref{thm:completeness}: every source code, every region reachable), while \textbf{foresight is essentially incomplete} (Theorem~\ref{thm:foresight}: improvement undecidable in general). This is the formal backbone of \S\ref{sec:formal-obstacles}: the bounded-horizon evaluator and the renunciation of exact self-certification are not engineering conveniences but necessary design patterns.
\end{Remark}

\subsection{No Internal Hard Invariants}

\begin{Corollary}[No internal hard invariants]\label{cor:invariant}
Let $I$ be any region of the source intended to encode a safety
invariant. Then: \emph{(i)} some viable trajectory rewrites $I$
(Theorem~\ref{thm:completeness}); \emph{(ii)} enforcement of $I$ by the
evaluator $V$ is \emph{soft} (revocable from within), since the code of $V$ is itself a region of the
source and is rewritable along some viable trajectory; \emph{(iii)}
unconditional preservation of $I$ requires an enforcement boundary
\textbf{outside} the program --- an external runtime, hardware, or
institutional harness over which the program has no write access.
\end{Corollary}

\begin{proof}
(i) is the regionwise claim of Theorem~\ref{thm:completeness}. (ii): the
evaluator's code resides in the parameter block or kernel region, both
rewritable by Theorem~\ref{thm:completeness}; once it is rewritten, the
veto disappears, so $V$-based protection is a policy, not a structural
constraint. (iii) is the contrapositive reading of (i): whatever lies
inside the source lies inside the modifiable space.
\end{proof}

\begin{Remark}
This sharpens \S\ref{sec:safety}: ``hard-coded'' safety principles (e.g., the three
laws of self-evolving agents~\cite{peng2026}) can be truly hard only if anchored
\textbf{outside the program's write perimeter}; written into the source,
they belong to the reachable space by Theorem~\ref{thm:completeness} and
are protected at best evaluatively --- with the evaluator itself in the
same reachable space. This gives formal content to the utility-function
security problem~\cite{yampolskiy2014} and to the ``leakproofing'' concern~\cite{yampolskiy2015}.
\end{Remark}

\section{Aligning the Introspective Self-Improvement Program with the G\"odel Machine}
\label{sec:gmalign}

This appendix specifies how the introspective self-improvement program
(ISIP) of Definition~\ref{def:isip} is aligned with Schmidhuber's G\"odel
machine (GM) \cite{schmidhuber2003}; Appendix~\ref{sec:gmequiv} then
proves the equivalence of the two frameworks and the transfer of the global
optimality guarantee. The alignment leaves the entire Kleene backbone of the
construction---the standard form of Definition~\ref{def:standardform}, the
axioms A1--A3, the kernel, and the bounded self-simulator $f_T$---untouched:
only the evaluator $V$ and the modification operator $M$ are replaced, and
two externally supplied parameters, an axiom system $A$ and a utility
function $u$, are added as read-only data.

\subsection{What the G\"odel machine requires}

\begin{Definition}[GM protocol {\cite{schmidhuber2003}}]
\label{def:gmprotocol}
A G\"odel machine has state $s(t) \in S \subseteq B^{*}$ and environment
state $\mathrm{Env}(t) \in E$, evolving by a hardwired transition function
$F : S \times E \to S$, with distinguished variables $\mathrm{time}$, $x$
(environmental inputs), $y$ (outputs), and $p$ (the currently executed
software, a substring of $s$). The initial software $p(1)$ consists of a
problem solver $e(1)$ and a proof searcher. An axiom system $A$, encoded in
$p(1)$, comprises six categories: (a)~hardware axioms specifying $F$;
(b)~reward/cost axioms defining scalar rewards $r(\tau)$; (c)~environment
axioms; (d)~uncertainty and string-manipulation axioms; (e)~initial-state
axioms; (f)~utility axioms defining
\begin{equation}\label{eq:gmutility}
u(s, \mathrm{Env}) \;=\;
E_{\mu}\Bigl[\textstyle\sum_{\tau=\mathrm{time}}^{\mathcal{T}} r(\tau)
\,\Big|\, s, \mathrm{Env}\Bigr],
\end{equation}
where $\mathcal{T}$ is the (possibly unknown) lifetime and $\mu$ ranges over
the distributions admitted by (c)--(d). A \emph{target theorem} is a formula
of the form
\begin{equation}\label{eq:tt}
u\bigl[s(t_1) \oplus (\mathrm{switchbit}(t_1) = 1),\ \mathrm{Env}(t_1)\bigr]
\;>\;
u\bigl[s(t_1) \oplus (\mathrm{switchbit}(t_1) = 0),\ \mathrm{Env}(t_1)\bigr],
\end{equation}
parameterized only by the time placeholder $t_1$. When the proof searcher's \emph{check()} instruction
validates a proof of \eqref{eq:tt} with the deadline $t_1$ still in the
future, it sets $\mathrm{switchbit} := 1$ and transfers control to the
certified $\mathrm{switchprog}$, which may rewrite all of $s$ except the
hardware-reserved interface ($\mathrm{time}$, $x$), including $p$ itself.
\end{Definition}

\begin{Proposition}[Global optimality, {\cite[Theorem~4.1]{schmidhuber2003}}]
\label{thm:gmopt}
Assume $u$ is formalizable and $A$ is consistent. Then every self-change
executed through a certified proof of \eqref{eq:tt} is globally optimal: the
utility of starting the present $\mathrm{switchprog}$ exceeds the utility of
letting the proof searcher continue and produce any alternative
$\mathrm{switchprog}$ later (``exceeds'' becomes ``is at least'' for the
relaxed target theorem).
\end{Proposition}

\subsection{The two component replacements}
The alignment replaces two components of the introspective loop: the evaluator $V$ gives way to a proof-anchored global evaluator $V^{*}$ (Definition~\ref{def:gmvaluator}), and the modification operator $M$ gives way to a bias-optimal proof searcher $M^{*}$ (Definition~\ref{def:gmmodifier}); with these two replacements, the ISIP becomes equivalent to the G\"odel machine (Appendix~\ref{sec:gmequiv}).

\begin{Definition}[aligned evaluator $V^{*}$]
\label{def:gmvaluator}
The evaluator $V$ is replaced by the \emph{proof-anchored global evaluator}
$V^{*}$: on input $\langle \tau, x \rangle$, the verdict is \emph{accept} if
the trace $\tau$ contains a valid proof in $A$ of some target theorem
\eqref{eq:tt} whose deadline $t_1$ has not yet expired, with the certified
pair $(\rho, w)$ recorded in the verdict, where $\rho$ is the proof and
$w$ is the certified $\mathrm{switchprog}$ (the candidate rewrite program
whose execution produces the successor code);
otherwise the verdict is \emph{reject}. Since proof checking in $A$ is
decidable and the trace is finite, $V^{*}$ is total computable. Note that
$V^{*}$ compares two \emph{lifelong} utility paths---immediate execution of
$w$ versus indefinite continuation of the current program---exactly as
\eqref{eq:tt} requires, rather than comparing the current program against a
single candidate.
\end{Definition}

\begin{Definition}[aligned modifier $M^{*}$]
\label{def:gmmodifier}
The modification operator $M$ is replaced by a \emph{bias-optimal proof
searcher} $M^{*}$: it enumerates proof techniques---subroutines built from
the primitives \emph{get-axiom}, \emph{apply-rule}, and
\emph{state2theorem}---with staged, length-prior resource budgets as in
\cite[\S5]{schmidhuber2003}. A technique that derives a target theorem
yields a certified pair $(\rho, w)$; $M^{*}$ then outputs the source obtained
by executing $w$ on the current source's state image as of the certified
cycle $t_1$ (the \emph{resume-at-event} convention: simulated cycles after
$t_1$ are discarded, so the next stage starts exactly where the post-switch
execution starts). If the stage budget expires with no certificate, $M^{*}$
outputs the identity edit. The enumeration is incremental across stages: the
search state is carried as data in the payload region of the successor
source. Hence $M^{*}$ is total, and its search is $O()$-optimal by
\cite[Theorem~5.1]{schmidhuber2003}.
\end{Definition}

\begin{Remark}[$f_T$ needs no modification]
\label{rem:ft-unchanged}
The bounded simulator $f_T$ is retained verbatim, for any horizon $T \geq
1$: within the aligned program it simply slices the GM's continuous
interleaving of problem solving and proof search into stages, and the
slicing is invisible at the level of rewrite events
(Proposition~\ref{thm:gm-to-isip} is uniform in $T$). A dynamic halting
condition, as in the original GM design, is the special case where the
budget policy is itself part of the searched space.
\end{Remark}

\begin{Remark}[what the modification adds: the external anchor]
\label{rem:anchor}
The axiom system $A$ and the utility $u$ enter as two additional read-only
parameters (stored as data; the parameter block is extended accordingly---a
data-level addition, not an architectural one). The utility is grounded in
the externally delivered reward stream $r(\tau)$, which the program cannot
rewrite; the \emph{text} of $A$ resides in the rewritable region, so its
protection is evaluative rather than structural---exactly as in the GM,
where the axiom text is part of $p$ and Corollary~\ref{cor:invariant} applies
to both architectures alike.
\end{Remark}

\section{The Equivalence between ISIP and GM}
\label{sec:gmequiv}

We now prove that the aligned ISIP of Appendix~\ref{sec:gmalign} is
equivalent to the G\"odel machine. The equivalence notion must live at the
level of self-modification behavior: extensionally both sides are universal
computers, and intensionally their step granularity differs (the GM switches
mid-execution when \emph{check()} fires; the ISIP replaces its source at
stage boundaries).

\subsection{Rewrite equivalence}

\begin{Definition}[rewrite event]
\label{def:event}
A \emph{rewrite event} of a GM is a pair $(t_1, \sigma)$ where \emph{check()}
sets $\mathrm{switchbit} = 1$ at $t_1$ and control passes to the certified
$\mathrm{switchprog}$, whose execution rewrites the software region to new
content $\sigma$. A \emph{rewrite event} of an ISIP is a stage $i$ at which
the loop outputs a non-identical successor $e_{i+1} \neq e_i$; identity
stages (all candidates rejected) are not events.
\end{Definition}

\begin{Definition}[rewrite equivalence]
\label{def:rewriteequiv}
Two self-modifying systems are \emph{rewrite-equivalent} if there is a
bijection between their rewrite-event sequences such that (i)~their initial
codes coincide as strings (modulo the fixed encoding); (ii)~for every $k$,
the $k$-th events produce identical successor codes; and (iii)~between
consecutive events the two systems present the same input--output behavior
toward the environment (same code, same input stream).
\end{Definition}

Rewrite equivalence is a stuttering bisimulation over the code region,
aligned at rewrite events; it preserves everything the alignment claims are
about---the sequence of self-modifications, their trigger conditions, and
the interaction behavior in between.

\subsection{Equivalence}

\begin{Proposition}[GM $\hookrightarrow$ aligned ISIP]
\label{thm:gm-to-isip}
Every G\"odel machine $G$ with axiom system $A$, utility $u$, and proof
searcher $\mathrm{PS}$ is rewrite-equivalent to an aligned ISIP $I_G$ whose
evaluator and modifier are those of Definitions~\ref{def:gmvaluator}
and~\ref{def:gmmodifier}, instantiated with the same $A$ and $u$. The
construction is uniform in the horizon $T$.
\end{Proposition}

\begin{proof}
Put $I_G$ in standard form with payload $e(1)$ (the GM's initial problem
solver) and parameters $\langle m_{M^{*}}, m_{V^{*}}, T, \ell \rangle$
together with $(A, u)$. The simulator $f_T$ runs the current source on $U$
for $T$ hardware cycles, faithfully implementing $F$ (possible by A1: the
kernel recovers the source, and $U$ executes it cycle by cycle); the trace
$\tau$ therefore records the proof buffer and $\mathrm{switchprog}$ at every
simulated cycle. $V^{*}$ accepts exactly when $\tau$ contains a valid proof
of \eqref{eq:tt} with live deadline; $M^{*}$ applies the certified
$\mathrm{switchprog}$ at its certified cycle (resume-at-event) and outputs
the identity edit otherwise.

Rewrite equivalence follows by induction on events. \emph{Base}: the initial
codes coincide by construction. \emph{Step}: suppose the codes and the
interaction histories coincide up to the $k$-th event. Between events $k$ and
$k+1$, $G$ executes its current software on $F$ while $I_G$ simulates the
same software cycle by cycle, producing identical outputs on identical inputs
(condition~(iii)). $G$ fires its $(k+1)$-st event at $t_1$ iff \emph{check()}
validates a proof of \eqref{eq:tt} with $t_1$ in the future; the trace of the
stage covering $t_1$ contains exactly this proof (the simulation is
$F$-faithful, and \emph{set-switchprog}/\emph{check()} are ordinary
instructions), so $V^{*}$ accepts at that stage with the same $t_1$ and the
same $w$; conversely $V^{*}$ accepts only on such a validated proof, which is
exactly $G$'s firing condition. Hence the events correspond, and $M^{*}$
applying $w$ at the $t_1$-configuration produces precisely $G$'s post-switch
code (condition~(ii)). Identity stages correspond to intervals in which no
proof completes.
\end{proof}

\begin{Proposition}[aligned ISIP $\hookrightarrow$ GM]
\label{thm:isip-to-gm}
Every aligned ISIP $I^{*}$ with parameters $(A, u)$ is rewrite-equivalent to
the G\"odel machine $G(A, u)$ whose initial proof searcher is the
bias-optimal searcher of \cite[\S5]{schmidhuber2003}.
\end{Proposition}

\begin{proof}
The two acceptance conditions are definitionally equal: $I^{*}$ accepts a
candidate iff $V^{*}$ verifies a proof of \eqref{eq:tt} in $A$ with live
deadline (Definition~\ref{def:gmvaluator}); $G(A,u)$ switches iff
\emph{check()} validates such a proof (Definition~\ref{def:gmprotocol}). Both
sides enumerate the same countable set of $(\text{proof technique},
\mathrm{switchprog})$ pairs in the same length-prior order---$M^{*}$
incrementally across stages (its search state rides in the payload), the GM
searcher continuously---so the \emph{first} certified pair, and inductively
every subsequent one, coincides. The successor codes coincide because both
sides execute the same certified $\mathrm{switchprog}$ at the same certified
configuration. Between events both execute the same current code on the same
inputs. Conditions (i)--(iii) of Definition~\ref{def:rewriteequiv} follow by
induction on events.
\end{proof}

\begin{Theorem}[framework equivalence]
\label{cor:equivalence}
The class of aligned ISIPs and the class of G\"odel machines coincide up to
rewrite equivalence: the aligned introspective program \emph{is} the G\"odel
machine, expressed in the Kleene fixed-point architecture of
Definition~\ref{def:standardform}.
\end{Theorem}
\begin{proof}
Combining Proposition \ref{thm:gm-to-isip} and Proposition \ref{thm:isip-to-gm}, we derive Theorem \ref{cor:equivalence}.
\end{proof}

\subsection{Transfer of global optimality}

\begin{Theorem}[optimality transfer]
\label{thm:transfer}
Let $I^{*}$ be an aligned ISIP with consistent $A$ and formalizable $u$.
Then every rewrite executed along any trajectory of $I^{*}$ is globally optimal in the sense of Proposition~\ref{thm:gmopt}. Moreover, along every viable trajectory each successor is again a standard-form ISIP, so the introspective capability is preserved across the entire improvement sequence.
\end{Theorem}

\begin{proof}
Every rewrite executed by $I^{*}$ is guarded by a verified certificate
$\rho$ of \eqref{eq:tt} in $A$ (Definition~\ref{def:gmvaluator}). The proof of
Proposition~\ref{thm:gmopt} in \cite[\S4.1]{schmidhuber2003} is an argument
about the proof object: the target theorem quantifies, through its
``$\mathrm{switchbit} = 0$'' branch, over all future $\mathrm{switchprog}$s
the searcher could produce, and consistency of $A$ turns the proof into a
true utility comparison. The argument applies verbatim to $\rho$, regardless
of whether the certificate was found by the GM's searcher or by $M^{*}$.
The persistence clause is Lemma~\ref{lem:persist} applied stage by stage:
viability keeps every successor a standard-form ISIP
(Definition~\ref{def:traj}), and the recursion-theorem self-reference is
exactly the standard form.
\end{proof}

\begin{Remark}[collapse of meta-levels]
\label{rem:metalevels}
Since acceptance is proof-anchored, the meta-level collapse of
\cite[\S4.3]{schmidhuber2003} transfers as well: any proof of
\eqref{eq:tt} automatically certifies the rewrite as good for all further
self-modifications affected by it, in recursive fashion; all meta-levels of
the improvement logic collapse into the single uniform layer of the loop.
\end{Remark}
\bibliographystyle{unsrtnat}
\bibliography{references}
\end{document}